\documentclass[10pt,a4paper]{article}

\usepackage[english]{babel}
\usepackage{graphicx}
\usepackage{hyperref} 
\usepackage{geometry}
\usepackage{amsmath}
\usepackage{amssymb}
\usepackage{amsfonts}
\usepackage{multicol}
\usepackage{slashed}
\usepackage{float}
\usepackage{caption}
\usepackage{subcaption}
\usepackage{cite}
\usepackage{amsthm}
\usepackage{xcolor}
\usepackage{musicography}
\numberwithin{equation}{section}

\newtheorem*{mydef}{Definition}

\newcommand{\M}{\mathcal{M}}
\newcommand{\R}{\mathbb{R}}
\newcommand{\E}{\mathbb{E}}
\newcommand{\Ls}{\mathcal{L}}

\newcommand{\p}{\partial}

\newcommand{\indep}{\perp \!\!\! \perp}

\def\Xint#1{\mathchoice
	{\XXint\displaystyle\textstyle{#1}}%
	{\XXint\textstyle\scriptstyle{#1}}%
	{\XXint\scriptstyle\scriptscriptstyle{#1}}%
	{\XXint\scriptscriptstyle\scriptscriptstyle{#1}}%
	\!\int}
\def\XXint#1#2#3{{\setbox0=\hbox{$#1{#2#3}{\int}$ }
		\vcenter{\hbox{$#2#3$ }}\kern-.6\wd0}}

\def\dashint{\Xint-}

\def\upint{\mkern13mu\overline{\vphantom{\intop}\mkern10mu}\mkern-20mu\int}
\def\lowint{\mkern3mu\underline{\vphantom{\intop}\mkern10mu}\mkern-10mu\int}

\title{\bf Stochastic Quantization on Lorentzian Manifolds}
\author{Folkert~Kuipers$^1$\thanks{E-mail: F.Kuipers@sussex.ac.uk}\\
	$^1${\em Department of Physics and Astronomy, University of Sussex,}\\{\em Brighton, BN1 9QH, United Kingdom}
}

\begin{document}

\maketitle

\begin{abstract}
We embed Nelson's theory of stochastic quantization in the Schwartz-Meyer second order geometry framework. The result is a non-perturbative theory of quantum mechanics on (pseudo-)Riemannian manifolds. Within this approach, we derive stochastic differential equations for massive spin-0 test particles charged under scalar potentials, vector potentials and gravity. Furthermore, we derive the associated Schr\"odinger equation. The resulting equations show that massive scalar particles must be conformally coupled to gravity in a theory of quantum gravity. We conclude with a discussion of some prospects of the stochastic framework.
\end{abstract}

\thispagestyle{empty}
\clearpage
\setcounter{page}{1}
\tableofcontents

\clearpage

\section{Introduction}

The construction of a theory of quantum gravity is one of the main open issues in theoretical high energy physics. One of the reasons why such a theory is desirable is that general relativity is unable to completely describe physical aspects of gravity at extremely high energy scales. This feature is most prominent in the fact that singularities seem to be unavoidable in general relativity, when natural assumptions are made~\cite{Schwarzschild:1916uq,Oppenheimer:1939ue,Penrose:1964wq,Hawking:1966vg}.
\par

From a physical perspective, the formation of such singularities would require the continuous collapse of a matter distribution to a delta distribution located at the singularity. On $\mathbb{R}^n$ one can make sense of such a collapse, as one can construct a family of smooth distributions that converges to the delta distribution. In general relativity, on the other hand, point-like sources cannot be obtained as a continuous limit of matter distributions defined on manifolds with smooth metrics, as the Einstein equations must be satisfied during the collapse \cite{Geroch:1987qn}.
\par 

It is expected that this paradox will be resolved, when general relativity is embedded into a quantum theory such that gravity is quantized. However, when one attempts such an embedding using standard quantum field theory methods, one runs into the problem that the resulting quantum theory is non-renormalizable~\cite{tHooft:1974toh}.
Up to the Planck scale, one can still make predictions regarding quantum gravity using effective field theory methods, since the ultra-violet divergences responsible for the non-renormalizability of the theory can be kept under control perturbatively. However, beyond the Planck scale this is no longer true, which renders the theory incomplete. 
\par 

Over the last decades many approaches to an ultra-violet complete theory of quantum gravity have been developed, and many interesting insights have been obtained within these approaches. In this paper, we argue that Nelson's stochastic quantization framework could help gain further insight in theories of quantum gravity. We will motivate this by showing that stochastic quantization allows to construct a well defined non-perturbative theory of quantum mechanics on (pseudo-)Riemannian manifolds.
\par 

We will adopt the framework of stochastic mechanics, also known as Nelsonian stochastic quantization\footnote{In this paper, we use the terms stochastic mechanics and stochastic quantization interchangeably. We emphasize that the framework is related to, but different from the Parisi-Wu formulation of stochastic quantization.}, that was proposed by F\'enyes~\cite{Fenyes} and Kershaw~\cite{Kershaw}, rederived by Nelson\cite{Nelson:1966sp,NelsonOld,Nelson} and further developed by many others. The main idea governing stochastic mechanics is that quantum mechanics can be derived from a stochastic theory. In this more fundamental theory all particles follow trajectories through a randomly fluctuating background field. Due to the interactions with this background field all matter behaves quantum mechanically. An equivalent way\footnote{One could call this a `passive' description of stochastic quantization, since the space-time fluctuates, while in the previous `active' description the matter defined on the space-time fluctuates.} to state this idea is that all particles and fields are defined on a randomly fluctuating space-time.
\par

We focus in this paper on ordinary quantum mechanics. We will thus work with point-like particles instead of fields. Moreover, we work on a fixed Lorentzian manifold. Therefore, the metric is not considered to be a dynamical field. We leave extensions to a field theory framework and to dynamical geometries for future work. In the stochastic quantization framework such extensions lead to a theory of quantum gravity.

\subsection{Stochastic quantization}

Since the quantization procedure in stochastic quantization is different from more commonly used quantization procedures, we will compare the main steps to canonical quantization. In a canonical quantization procedure one starts with a classical Hamiltonian $H(p,x)$ and promotes the variables $p,x$ to operators $P,X$ such that
\begin{equation*}
	H(p,x) \rightarrow \hat{H}(P,X).
\end{equation*}
One then imposes canonical commutation relations
\begin{equation}
	[X^\nu,P_\mu] = i \, \hbar\, \delta_\mu^\nu.
\end{equation}
Moreover, one postulates the existence of a wave function $\Psi$, which is an element of a complex Hilbert space with $L^2$-norm, that can be used to calculate observables, i.e.,
\begin{equation}
	\langle \Psi | \hat{O} |\Psi\rangle = O.
\end{equation}
\par 

In stochastic quantization, one starts with a classical Lagrangian $L_c(x,v,\tau)$, and promotes the position of a particle $x$ to a stochastic process $X(\tau)$. Since the stochastic process is not differentiable, one can define two velocities $v_\pm$ using conditional expectations:
\begin{align}
	v_+(X(\tau),\tau) &= \lim_{h\downarrow 0}\frac{1}{h} \E\left[X(\tau+h) - X(\tau) | X(\tau)\right],\nonumber\\
	v_-(X(\tau),\tau) &= \lim_{h\downarrow 0} \frac{1}{h} \E\left[X(\tau) - X(\tau-h) | X(\tau)\right].
\end{align}
One can then introduce a stochastic Lagrangian
\begin{equation}
	L_c(x,v,\tau) \rightarrow L(X,V_+,V_-,\tau) = \frac{1}{2}\left[ L_c(X,V_+,\tau) + L_c(X,V_-,\tau) \right]
\end{equation}
Moreover, one fixes the quadratic variation\footnote{More commonly used notations for $d[[X^i,X^j]]$ are $d[X^i,X^j]$ or $dX^idX^j$. We use the double brackets instead to avoid confusion with the commutator, first order bilinear tensors and second order vectors that will be introduced in section \ref{sec:Geometry}.} of the process $X$ by the background hypothesis:
\begin{equation}\label{eq:BackgroundHyp}
	[[X^\nu,X_\mu]](\tau) = \frac{\hbar}{m}\, \delta_\mu^\nu\, \tau.
\end{equation}
We remind the reader that the joint quadratic variation of two processes $X,Y$ is itself a stochastic process and can be written as
\begin{equation}
	[[X,Y]](\tau) = X(\tau) Y(\tau) - X(0) Y(0) - \lowint_{0}^\tau X(s) dY(s) - \lowint_{0}^\tau Y(s) dX(s).
\end{equation}
The It\^o integral used in this expression is defined by
\begin{equation}
	\lowint_{\tau_i}^{\tau_f} f(X,\tau)\, d X
	:=
	\lim_{k\rightarrow\infty} \sum_{[\tau_{j},\tau_{j+1}]\in \pi_k} f (X(\tau_j),\tau_j) \left[ X(\tau_{j+1}) - X(\tau_{j})\right],
\end{equation}
where $\pi_k$ is a partition of $[\tau_i,\tau_f]$. 
\par

Observables in stochastic quantization can be calculated using the expectation $\E$, which is defined on a filtered probability space, and evaluated as a Lebesgue integral in the $L^2$-space of stochastic processes. The construction of expectation values in modern probability theory as founded by Kolmogorov \cite{Kolmogorov} requires the existence of a probability measure $\mathbb{P}$ in the probability space, and a measure $\mu$ in the $L^2$-space, but not the existence of a probability density.\footnote{If a probability density $\rho(x)$ exists, one has the familiar relation $d\mu(x)=\rho(x)d^n x$.} Therefore, the wave function $\Psi$ no longer needs to be postulated in stochastic quantization. 
\par

Since the wave function is no longer fundamental to the theory, the interpretation of quantum mechanics in the stochastic quantization framework is different from the standard Copenhagen interpretation. In stochastic quantization, one assumes that particles follow well defined trajectories through space-time. However it is assumed that all matter moves through a fluctuating background field, which is sometimes called the aether, but can also be regarded as a fluctuating space-time or as a diffeomorphism invariant quantum vacuum.
\par

Due to the fluctuating background field, the motion of massive particles\footnote{Stochastic quantization has yet to be extended to massless particles.} will become stochastic and comparable to a frictionless Brownian motion.\footnote{Notice that eq.~\eqref{eq:BackgroundHyp} characterizes a scaled Brownian motion~\cite{Levy}.} This Brownian motion is imposed to be time-reversible. This additional assumption introduces an important distinction from Brownian motion processes that are more commonly studied in statistical physics.
\par

Most stochastic diffusion processes that are studied in physics, such as for example the Ornstein-Uhlenbeck process, are dissipative diffusions. These processes are not time reversible, and energy is transferred from the system to the environment until an equilibrium is reached. The processes studied in stochastic mechanics are conservative diffusion processes. These processes are time-reversible and the expected energy transfer between the system and environment is $0$ at all times. 
\par

The fact that the wave function is no longer fundamental in stochastic quantization has two further important consequences. First, constructing normalized wave functions on Riemannian manifolds is a difficult task, that complicates extensions of ordinary quantum mechanics to manifolds. This problem is circumvented in the stochastic approach, as the wave function no longer needs to exist globally.
\par 

Secondly, due to the secondary role of the wave function, there is no measurement problem in stochastic mechanics. The wave function and probability density in stochastic mechanics have the same status as in standard probability theory. A theoretically perfect measurement in stochastic mechanics thus corresponds to conditioning of the process. Conditioning is a mathematical operation that still leads to collapse of the wave function, but since the wave function is only a mathematical construct and not a physical object, this does not correspond to a physical interaction.

\subsection{Successes of stochastic quantization}
The success of stochastic quantization relies on the relation between probability density functions associated to stochastic processes and partial differential equations. In the case of dissipative diffusions, the probability density associated to the solution of a stochastic differential equation evolves according to a parabolic differential equation. This result is known as the Feynman-Kac formula \cite{FKac}. An example of this relation is the fact that the probability density of a dissipative Brownian motion evolves according to the heat equation, which is a real diffusion equation.
\par

A similar relation exists for conservative diffusion processes. For example, the probability density of a conservative Brownian motion evolves according to the Schr\"odinger equation, which is a complex diffusion equation. This result is closely related to the Feynman-It\^o formula \cite{FIto,Albeverio}. 
Before this latter relation was formally established, it was discovered independently by F\'enyes, Kershaw and Nelson \cite{Fenyes,Kershaw,Nelson:1966sp,NelsonOld,Nelson} that the Schr\"odinger equation can be derived from a stochastic theory, if one assumes that particles follow a time-reversible stochastic process, governed by a stochastic version of Newton's second law, where the force is derived from a potential.

\par
The theory that was developed in this way is called stochastic mechanics. The immediate consequence of this discovery is that all predictions of quantum mechanics that follow from the Schr\"odinger equation, are also predictions of stochastic mechanics.
Later it was shown that the same result can be formulated in terms of Lagrangian dynamics using the stochastic variational calculus developed by Yasue~\cite{Yasue,Yasue:1981wu,Zambrini}. This Lagrangian approach goes by the name of stochastic quantization.
\par

The theory of stochastic mechanics and stochastic quantization has been extended to Riemannian manifolds, see e.g. Refs.~\cite{Dankel,DohrnGuerraI,DohrnGuerraII,Dohrn:1985iu,Guerra:1982fn,Nelson}. Moreover, extensions of stochastic quantization to bosonic field theory have been developed, cf. e.g. Refs.~\cite{Guerra:1973ck,GuerraRuggiero,Guerra:1980sa,Guerra:1981ie,Kodama:2014dba,Marra:1989bi,Morato:1995ty,Garbaczewski:1995fr,Pavon:2001}. Furthermore, the notion of spin has been discussed in this framework, cf. e.g. Refs.~\cite{Nelson,Dankel,Fritsche:2009xu}. 
\par 

It is worth noticing that in the dissipative field theoretic stochastic framework that was later developed by Parisi and Wu \cite{Parisi:1980ys,Damgaard:1987rr}, and also goes by the name of stochastic quantization, extensions to fermionic field theories have been developed, cf. e.g. Ref.~\cite{Damgaard:1983tq}. Although this framework is different from the stochastic quantization as developed by Nelson and others, there exist many similarities. It is also worth mentioning that several authors have incorporated stochastic mechanics into models of quantum gravity, cf. e.g. Refs.~\cite{Markopoulou:2003ps,Erlich:2018qfc}.
\par

Many basic results from quantum mechanics such as the commutation relations, the uncertainty principle, the double slit experiment and the motion of particles in various potentials have been discussed within the stochastic framework, see e.g. Refs.~\cite{Zambrini,Nelson,Guerra:1981ie,Pena,Olavo,Petroni_2000,Gaeta}. We emphasize that the interpretation of these results radically changes in the stochastic quantization framework, as the particle follows a well defined trajectory. For example, in the double slit experiment, a particle always goes through one slit. One still obtains an interference pattern, as this is the unique solution of the time-reversible diffusion process.\footnote{Let us be a bit more precise, as the process is slightly more complicated in stochastic quantization: after passing through one of the slits, the particle will diffuse according to a one slit diffusion process. However, due to the imposed time-reversibility of the motion, it will transition into a double slit diffusion process. The length scale associated to this transition is the width of the slit, cf. e.g. sections 16 and 17 in Ref.~\cite{Nelson}.}

\subsection{Criticism on stochastic quantization}
Despite the successes described above, stochastic quantization has never been widely studied. We will therefore review some of the main concerns that have been raised against stochastic quantization.
\par 

Historically, one of the more prominent confusions arose from the idea that a diffusion process is necessarily dissipative, and cannot give rise to quantum mechanics. As argued before, this is not the case, when the diffusion is time-reversible. This point has been well explained by Nelson in section 14 of Ref.~\cite{Nelson}, where an analogy is made with the difference between Aristotelean and Galilean dynamics. It should be noted that in order to describe entanglement in stochastic quantization, the background field has to be non-local. This particular feature was disliked by Nelson, cf. e.g. Ref.~\cite{Nelson_2012}. We stress that this non-locality is merely a feature of quantum mechanics, and not specific to stochastic quantization. Moreover, it is an open question, whether the non-locality of the background can be avoided, if one considers non-Markovian diffusion processes.
\par

Another concern that may be raised against stochastic mechanics is that it can be regarded as a hidden variable theory, as it is assumed that a background field exists that is responsible for the quantum fluctuations. One could thus expect that stochastic mechanics satisfies the Bell inequalities, which would distinguish it from quantum mechanics. We will avoid this issue by assuming that the background field is fundamentally random, in the sense that the fluctuations cannot be derived from a more fundamental theory. Under this assumption there are no deterministic hidden variables. This assumption distinguishes the framework from for example the Brownian motion of a colloid suspended in a liquid, where the trajectory of the colloid can in theory be derived by solving the equations of motion of all the molecules in the liquid.
\par 

A more pressing issue for stochastic quantization is Wallstrom's criticism\cite{Wallstrom:1988zf,WallstromII}, which states that the $2\pi$ periodicity of the wave function has to be imposed as an additional assumption. Such an assumption must be made ad hoc, since the wave function is not a fundamental object in the theory. Several responses against this criticism have been given, such as for example the incorporation of zitterbewegung \cite{Derakhshani:2015cda,Derakhshani:2016diw}, adding a postulate regarding the boundedness of the Laplace operator acting on the probability density \cite{Schmelzer} or by adding the assumption of unitarity of superpositions of wave functions \cite{Fritsche:2009xu}. It is also worth mentioning that it was pointed out in Ref.~[11] that the stochastic processes should be lifted to the universal cover of the configuration space, as the configuration space itself might not be simply connected. When this is done, the wave function obtains periodicity factors that are related to the winding numbers around the holes in the configuration space, which could resolve Wallstrom's criticism.
\par

Since no consensus yet exists about the solution of Wallstrom's criticism, we will take a more pragmatic approach: we accept this ad hoc constraint and remain agnostic about its solution. The reason for this is that imposing such a constraint is only problematic at a foundational level. Even if Wallstrom's criticism cannot be resolved within stochastic quantization, the theory can still be used as an alternative mathematical model of quantum theory, and can thus be used to make predictions about quantum systems. As we will show in this paper, a particular advantage of the stochastic model is that it can be formulated on (pseudo-)Riemannian manifolds, which could help guide the way towards a theory of quantum gravity. 
\par

A more practical concern regarding stochastic quantization is that analytical calculations require to solve stochastic differential equations. This is notoriously difficult. In fact, an important solution method relies on the mapping stochastic differential equations to path integral problems and to partial differential equations, as established by the Feynman-Kac formula. It is thus expected that many calculations can more easily be performed in ordinary quantum theory. This would render stochastic mechanics as an alternative mathematical model unnecessary. Despite this fact, it is expected that stochastic quantization could prove to be useful in numerical calculations, and a small number of analytical calculations. More interesting, however, is the potential of stochastic quantization on a more formal level. In particular, it could prove to be useful in mathematically rigorous definitions of the path integral, which is expected to be essential for constructing a theory of quantum gravity. We note here that stochastic approaches already serve as one of the stepping stones of the Euclidean approach in quantum field theory, see e.g. \cite{Wick:1954eu,Schwinger:1958mma,Symanzik,NelsonPath}.

\subsection{Postulates of the theory}
Before moving on, let us summarize the fundamental assumptions of stochastic quantization: we assume that all particles follow well defined trajectories through a diffeomorphism invariant background field. This background field induces stochastic fluctuations such that the motion of particles resembles a conservative Brownian motion. Moreover, the quadratic variation of this process scales with the Planck constant according to the background hypothesis. We have the following postulates:
\begin{itemize}
	\item All observables are invariant under a change of coordinate system.
	\item The stochastic motion of a particle with mass $m$ is Markovian.
	\item The stochastic motion of a particle with mass $m$ is time-reversible.
	\item The stochastic motion obeys the structure equation $[[X_\mu,X^\nu]](\tau)=\frac{\hbar}{m}\delta_\mu^\nu\tau$.
\end{itemize}
We note that the classical limit of the theory can be obtained straightforwardly by taking the limit $\hbar\rightarrow0$.

\subsection{Main results of the paper}
In this paper, we work in the $(-+++)$ signature with a Riemann tensor defined by $\mathcal{R}^\rho_{\;\;\sigma\mu\nu} = \p_\mu \Gamma^\rho_{\nu\sigma} - \p_\nu \Gamma^\rho_{\mu\sigma} + \Gamma^\rho_{\mu\kappa} \Gamma^\kappa_{\nu\sigma} - \Gamma^\rho_{\nu\kappa} \Gamma^\kappa_{\mu\sigma}$ and Ricci tensor $\mathcal{R}_{\mu\nu}= \mathcal{R}^\rho_{\;\;\mu\rho\nu}$. In addition, we set $c=1$ throughout the paper.
\par 

The main result we present in this paper is the following: in the stochastic quantization framework, a massive scalar particle moving on a Lorentzian manifold and governed by the stochastic Lagrangian 
\begin{equation}
	L(X,V_+,V_-,\tau) 
	= 
	\frac{1}{2} L_c(X,V_+,\tau) + \frac{1}{2} L_c(X,V_-,\tau)
\end{equation}
where the classical Lagrangian is given by
\begin{equation}\label{eq:ClassLagrangianIntro}
	L_c(x,v,\tau) 
	= 
	\frac{m}{2}\, g_{\mu\nu}(x)\, v^\mu\, v^{\nu}
	- \hbar\, A_\mu(x,\tau)\, v^\mu
	- \mathfrak{U}(x,\tau)
\end{equation}
with $x=(t,\vec{x})$ and $\tau$ is the proper time, evolves according to the Stratonovich stochastic differential equation 
\begin{align}\label{eq:EoMIntro}
	m \, g_{\mu\nu} \left( 
	d^2 X^\nu 
	+ \Gamma^\nu_{\rho\sigma} \, dX^\rho dX^\sigma 
	\right)
	&=
	\left(
	\hbar\, \p_\tau A_\mu - \nabla_\mu \mathfrak{U} - \frac{\hbar^2}{12 m} \nabla_\mu \mathcal{R}  
	\right) d\tau^2 \nonumber\\
	&\quad
	- \hbar \left(
	\nabla_\mu A_\nu - \nabla_\nu A_\mu 
	\right) dX^\nu d\tau.
\end{align}
Furthermore, if the probability density $\rho(x,\tau)$ associated to the probability measure $\mu=\mathbb{P} \circ X^{-1}$ exists, one can construct the wave function
\begin{equation}\label{eq:WaveFunction}
	 \Psi(x,\tau) = \sqrt{\rho(x,\tau)} \, \exp \left\{ \frac{i}{\hbar}\, \E \left[ \int_{\tau_i}^{\tau} L\big(X(t),V_+(t),V_-(t),t\big) \, dt \Big| X(\tau) = x \right] \right\}	
\end{equation}
that evolves according to a generalization of the Schr\"odinger equation given by 
\begin{equation}\label{eq:SchrodingerIntro}
	i \hbar\, \frac{\p}{\p\tau}\Psi 
	=
	\left[ 
	- \frac{\hbar^2}{2m}\left(
	\Big[\nabla_\mu + i A_\mu \Big] \Big[\nabla^\mu + i A^\mu \Big] 
	- \frac{1}{6} \mathcal{R} \right) + \mathfrak{U} 
	\right] \Psi.
\end{equation}
This wave function obeys the Born rule
\begin{equation}\label{eq:BornIntro}
	|\Psi(x,\tau)|^2 = \rho(x,\tau).
\end{equation}
If there is no explicit proper time dependence in $A_\mu$ or $\mathfrak{U}$, one can solve by separation of variables such that
\begin{equation}
	\Psi(x,\tau) = \sum_{k} \phi_k(x) \exp\left(\frac{i\, m \, \lambda_k}{2\, \hbar} \tau \right),
\end{equation}
where $\phi_k(x)$ solves the generalization of the Klein-Gordon equation given by
\begin{equation}
	\hbar^2 \left(
	\Big[\nabla_\mu + i A_\mu\Big] 
	\Big[\nabla^\mu + i A^\mu\Big] 
	- \frac{1}{6} \mathcal{R} 
	\right) \phi_k 
	= 
	m^2\, \lambda_k\, \phi_k  
	+ 2\,m\, \mathfrak{U}\, \phi_k.
\end{equation}
\par

We note that the derivation of eqs.~\eqref{eq:WaveFunction},~\eqref{eq:SchrodingerIntro}~and~\eqref{eq:BornIntro} is a well established result on $\R^n$, see e.g. Refs.~\cite{Kershaw,Nelson:1966sp,NelsonOld,Nelson,Zambrini,Guerra:1981ie}. Moreover, partial extensions to Riemannian manifolds have been known for some time, cf. Refs.~\cite{Dankel,DohrnGuerraI,DohrnGuerraII,Dohrn:1985iu,Guerra:1982fn,Nelson}.
\par

In this paper, we show that these results can be generalized to pseudo-Riemannian manifolds. An important ingredient for these extensions is the second order geometry as developed by Schwartz and Meyer \cite{Schwartz,Meyer,Emery}. This is an extension of ordinary differential geometry that allows to describe stochastic processes on manifolds. In addition to the extension of stochastic quantization to pseudo-Riemannian manifolds, we will give some new interpretations of stochastic quantization.
\par

This paper is organized as follows: in the next section, we review second order geometry; in section 3, we introduce the relevant semi-martingale processes for quantum mechanics; section 4 discusses integration along semi-martingales on manifolds; in section 5, we discuss stochastic variational calculus; in section 6, we discuss the shape of the stochastic action; in section 7, we put everything together and derive the stochastic differential equations for quantum mechanical scalar test particles on pseudo-Riemannian manifolds, and the associated Schr\"{o}dinger equation. Finally, in section 8, we conclude and summarize some future perspectives of the stochastic approach.

\section{Second Order Geometry}\label{sec:Geometry}
In this section, we review the theory of Schwartz-Meyer second order geometry, that can be used to extend the theory of stochastic calculus to manifolds. The first three subsections are loosely based on Ref.~\cite{Emery}. The later subsections contain new material and extend some important concepts from first order geometry into second order geometry. For more detail we refer to the work of Emery \cite{Emery} and the original works by Schwartz \cite{Schwartz} and Meyer \cite{Meyer}.

\subsection{Second order vectors and forms}\label{sec:SecVec}

We consider a $(n=d+1)$-dimensional pseudo-Riemannian manifold $\M$ with the usual first order tangent and cotangent spaces $T_x\M,\, T^\ast_x\M$. For every $x\in\M$ and any coordinate chart containing $x$ one can write down bases for the tangent and cotangent space respectively given by $\{\p_\mu|\mu\in\{0,1,...,d\}\}$ and $\{dx^\mu|\mu\in\{0,1,...,d\}\}$. In particular for $v\in T_x\M$ and $\omega\in T^\ast_x\M$ we have
\begin{align}
	v &=v^\mu \, \p_\mu,\nonumber\\
	\omega &=\omega_\mu \, dx^\mu.
\end{align}
Furthermore, a form $\omega\in T^\ast_x\M$ can often be written as the differential form of some function $f:\M\rightarrow\R$ i.e.
\begin{equation}
	\omega = df = \partial_\mu f \, dx^\mu.
\end{equation}
The product rule for such differential forms is given by
\begin{equation}
	d(fg) = f\, dg + g\, df.
\end{equation}
In addition, there exists a metric associated to the tangent space that is given by
\begin{equation}\label{Metric1}
	g: T_x\M \otimes T_x\M \rightarrow \R 
	\quad {\rm s.t.} \quad (v,w) 
	\mapsto \langle v | w \rangle = g_{\mu\nu} v^\mu w^\nu,
\end{equation}
and is bilinear, symmetric and non-degenerate. Moreover the metric induces an isomorphism $g^{\musFlat}:T_x\M\rightarrow T^\ast_x\M$ between the tangent and cotangent space, that is defined by
\begin{equation}
	\langle g^{\musFlat}(v),w \rangle = \langle v, g^{\musFlat}(w) \rangle = \langle v|w\rangle
\end{equation}
We define a similar bracket for two forms $\alpha,\beta\in T_x^\ast\M$ by
\begin{equation}
	\langle \alpha|\beta \rangle 
	= \langle \alpha,g^{\musSharp}(\beta) \rangle 
	= \langle g^{\musSharp}(\alpha),\beta\rangle.
\end{equation}
\par

We will now define a second order tangent space and cotangent space $\tilde{T}_x\M,\, \tilde{T}^\ast_x\M$. For every $x\in\M$ and any coordinate chart containing $x$ one can write down bases for the tangent and cotangent space respectively given by $\left\{\p_\mu, \p_{\mu\nu}\big|\mu\leq\nu\in\{0,1,...,d\}\right\}$ and $\{d_2x^\mu,dx^\mu \cdot dx^\nu| \mu\leq\nu\in\{0,1,...,d\}\}$.\footnote{Notice that $dx^\mu \cdot dx^\nu \neq dx^\mu \otimes dx^\mu$.} In particular, for $V\in \tilde{T}_x\M$ and $\Omega\in \tilde{T}^\ast_x\M$ we have
\begin{align}
	V &=v^\mu \, \p_\mu + v^{\mu\nu} \, \p_{\mu\nu},\nonumber\\
	\Omega &=\omega_\mu\, d_2x^\mu + \omega_{\mu\nu}\, dx^\mu\cdot dx^\nu.
\end{align}
Notice that $T_x\M\subset\tilde{T}_x\M$ and $T^\ast_x\M\subset\tilde{T}^\ast_x\M$. Furthermore, $\p_{\mu\nu}:=\p_\mu\p_\nu$ is a symmetric object, which implies that $v^{\mu\nu}$ must be symmetric. Moreover, we choose the basis of the cotangent space dual to the basis of the tangent space. This imposes $dx^\mu\cdot dx^\nu$, and $\omega_{\mu\nu}$ to be symmetric as well.
\par

We have a duality pairing between the bases of the tangent and cotangent space such that:
\begin{align}\label{DualPairing}
	\langle \p_\mu, d_2x^{\rho} \rangle &= \delta_\mu^\rho, \nonumber\\
	\langle \p_\mu, dx^{\rho} \cdot dx^{\sigma} \rangle &= 0, \nonumber\\
	\langle \p_{\mu\nu}, dx^{\rho} \rangle &= 0, \nonumber\\
	\langle \p_{\mu\nu}, dx^{\rho}  \cdot dx^{\sigma} \rangle &= \frac{1}{2}\left(\delta_\mu^\rho \, \delta_\nu^\sigma + \delta_\mu^\sigma \, \delta_\nu^\rho \right).
\end{align}
The duality pairing of an arbitrary vector and covector is then given by
\begin{equation}\label{eq:InProd}
	\langle V, \Omega \rangle = v^\mu \omega_\mu + v^{\mu\nu}\omega_{\mu\nu}.
\end{equation} 
As in the classical case, forms $\Omega\in\tilde{T}^\ast_x\M$ can often be written as a differential form of some function $f:\M\rightarrow \R$:
\begin{equation}
	\Omega = d_2 f = \p_\mu f\, d_2 x^\mu + \p_{\mu\nu} f \, dx^\mu\cdot dx^\nu.
\end{equation}
The product rule for differential forms is given by
\begin{equation}\label{ProductRule2}
	d_2 (f g) = f\, d_2 g + g\, d_2 f + 2\, df \cdot dg
\end{equation}
where the product of first order forms\footnote{More generally, one often defines the \textit{carr\'e du champ} operator or the \textit{squared field operator} associated to a linear mapping $L$ for two functions $f,g$ by $\Gamma(f,g):=\frac{1}{2} \left[L(fg) - f\,Lg- g\,L f\right]$. Cf. e.g. Lemma 6.1 in \cite{Emery}. We can then interpret $df\cdot dg$ as the squared field operator associated to the second order differential operator $d_2$ acting on $f,g$.} $\omega,\theta\in T_x\M$ is defined by
\begin{align}
	\omega \cdot \theta
	&:= \frac{1}{2} \left(
	\omega_\mu \, \theta_\nu + \omega_\nu\, \theta_\mu \right) 
	dx^\mu \cdot dx^\nu\nonumber\\
	&= \omega_\mu \, \theta_\nu \, dx^\mu \cdot dx^\nu.
\end{align}
Therefore, the product for two first order differential forms can be written as
\begin{equation}\label{SqFieldOP}
	df \cdot dg = \p_\mu f \, \p_\nu g \, dx^\mu \cdot dx^\nu.
\end{equation}
\par

It will be useful to define mappings between the first order and second order tangent spaces. The projection map\footnote{In Ref.~\cite{Emery} this map is called the restriction $R$.} can be defined as:
\begin{equation}\label{eq:ProjectionMap}
	\mathcal{P}: \tilde{T}_x^\ast\M \rightarrow T^\ast_x\M 
	\quad \textrm{s.t.} \quad 
	\begin{cases}
		\mathcal{P}(d_2f) = d f,\\
		\mathcal{P}(\omega \cdot \theta) = 0.
	\end{cases}
\end{equation}
Furthermore, there exists a unique smooth and invertible linear map $\mathcal{H}$ from bilinear first order forms to second order forms, such that $\mathcal{P}\circ \mathcal{H} = 0$, given by\footnote{cf. Proposition 6.13 in Ref.~\cite{Emery}.}
\begin{equation}
	\mathcal{H}: T^\ast_x\M \otimes T^\ast_x\M \rightarrow \tilde{T}^\ast_x\M
	\quad {\rm s.t.} \quad
	(\omega,\theta) \mapsto \omega \cdot \theta,
\end{equation}
The adjoint of this map is denoted by $\mathcal{H}^\ast:\tilde{T}_x\M \rightarrow T_x\M \otimes T_x\M$. In addition there exists a unique linear map\footnote{cf. Theorem 7.1 in Ref.~\cite{Emery}. We use an underlined $\underline{d}$ to avoid confusion with the exterior derivative.} $\underline{d}:T^\ast_x\M\rightarrow \tilde{T}^\ast_x\M$ such that for any $f\in C^\infty(\M,\R)$, $\omega\in T^\ast_x\M$ and $u,v\in T_x\M$
\begin{align}
	\underline{d}(df) &= d_2 f,\nonumber\\
	\underline{d}(f\omega) &= f \underline{d}\omega + df \cdot \omega,\nonumber\\
	\langle \underline{d}\omega, [u,v] \rangle &= \langle \omega, [[u , v]] \rangle,\nonumber\\
	\langle \underline{d}\omega, \{u,v\} \rangle &= u \langle \omega, v \rangle + v \langle \omega, u \rangle,
\end{align}
where $[u,v]$ is the commutator,  $\{u,v\}$ the anti-commutator and $[[u , v]]$ the joint quadratic variation of $u$ and $v$.
\par

Finally,\footnote{cf. Proposition 7.28 in Ref.~\cite{Emery}} one can define maps $\mathcal{F}:\tilde{T}_x\M\rightarrow T_x\M$ and $\mathcal{G}:T^\ast_x\M\rightarrow \tilde{T}^\ast_x\M$ such that for any affine connection\footnote{$\mathfrak{X}(\M)$ is the space of all smooth vector fields on $\M$, i.e. the space of all smooth sections of the tangent bundle $T\M$.} $\Gamma:\mathfrak{X}(\M) \times \mathfrak{X}(\M) \rightarrow \mathfrak{X}(\M)$ the following relations define a bijection between $\mathcal{F}$ and $\Gamma$
\begin{align}
	(\mathcal{F}\, V) f &= V\, f - \langle \mathcal{H} \, \Gamma^\ast(df), V \rangle,\nonumber\\
	\Gamma(u,v)f &= u\, v\, f - \mathcal{F}(u\, v) f,
\end{align}
where $V$ is a second order vector and $u,v$ are first order vector fields. A bijection between $\mathcal{G}$ and $\Gamma$ is then defined by
\begin{align}
	\mathcal{G}(df) &= d_2 f - \mathcal{H} \, \Gamma^\ast(df),\nonumber\\
	\Gamma(u,v)f &= u\,v\, f - \langle \mathcal{G}(df), u\,v \rangle.
\end{align}
Moreover, $\mathcal{F}$ and $\mathcal{G}$ are each others adjoint.\footnote{It is possible to take a connection in the defining relation for $\mathcal{G}$ that is different from $\mathcal{F}$. If such a choice is made, $\mathcal{F}$ and $\mathcal{G}$ are no longer each others adjoint. In this paper, we will not make such a choice, as we will restrict ourselves to the Levi-Civita connection.}

\subsection{Coordinate transformations}
In this section, we investigate the change of vectors and covectors under coordinate transformations. For a vector field $V$ we find:
\begin{align}
	V f &= \left(v^\mu \p_\mu + v^{\mu\nu} \p_{\mu\nu} \right)f \nonumber\\
	&= \left(v^\mu \frac{\p \tilde{x}^\rho}{\p x^\mu} \tilde{\p}_\rho
	+ v^{\mu\nu} \p_\mu\left[\frac{\p \tilde{x}^\rho}{\p x^\nu} \tilde{\p}_\rho \right] \right) f \nonumber\\
	&= \left(v^\mu \frac{\p \tilde{x}^\rho}{\p x^\mu} \tilde{\p}_\rho
	+ v^{\mu\nu} \frac{\p^2 \tilde{x}^\rho}{\p x^\mu \p x^\nu} \tilde{\p}_\rho
	+ v^{\mu\nu} \frac{\p \tilde{x}^\sigma}{\p x^\mu}  \frac{\p \tilde{x}^\rho}{\p x^\nu} \tilde{\p}_{\sigma\rho} \right) f.
\end{align}
Hence, we find the active transformations laws
\begin{align}
	v^\mu \rightarrow \tilde{v}^\mu 
	&= v^{\rho} \frac{\p \tilde{x}^\mu}{\p x^\rho} 
	+ v^{\rho\sigma} \frac{\p^2 \tilde{x}^\mu}{\p x^\rho \p x^\sigma},\nonumber\\
	v^{\mu\nu} \rightarrow \tilde{v}^{\mu\nu} 
	&= v^{\rho\sigma} \frac{\p \tilde{x}^\mu}{\p x^\rho} \frac{\p \tilde{x}^\nu}{\p x^\sigma},
\end{align}
or equivalently the passive transformation laws
\begin{align}
	\p_\mu \rightarrow \tilde{\p}_\mu &= \frac{\p x^\rho}{\p\tilde{x}^\mu} \p_\rho,\nonumber\\
	\p_{\mu\nu} \rightarrow \tilde{\p}_{\mu\nu}
	&= \frac{\p^2 x^\rho}{\p \tilde{x}^\mu \p \tilde{x}^\nu} \p_\rho
	+ \frac{\p x^\sigma}{\p \tilde{x}^\mu}  \frac{\p x^\rho}{\p \tilde{x}^\nu} \p_{\rho\sigma}.
\end{align}
A form $\Omega$ transforms as
\begin{align}
	\Omega(V f) &= \left(\omega_\mu d_2x^\mu + \omega_{\mu\nu} dx^\mu \cdot dx^\nu \right)(V f) \nonumber\\
	&= \left(
	\omega_\mu \frac{\p x^\mu}{\p \tilde{x}^\rho} d_2 \tilde{x}^\rho 
	+ 
	\omega_\mu \frac{\p^2 x^\mu}{\p \tilde{x}^\rho \p \tilde{x}^\sigma} 
	d\tilde{x}^\rho \cdot d\tilde{x}^\sigma
	+ 
	\omega_{\mu\nu} \frac{\p x^\mu}{\p \tilde{x}^\rho} 
	\frac{\p x^\nu}{\p \tilde{x}^\sigma}
	d\tilde{x}^\rho \cdot d\tilde{x}^\sigma 
	\right)(V f).
\end{align}
Therefore, the active transformation laws are given by
\begin{align}
	\omega_\mu \rightarrow \tilde{\omega}_\mu 
	&= \omega_{\rho} \frac{\p x^\rho}{\p \tilde{x}^\mu},\nonumber\\
	\omega_{\mu\nu} \rightarrow \tilde{\omega}_{\mu\nu} 
	&= \omega_{\rho} \frac{\p^2 x^\rho}{\p \tilde{x}^\mu \p \tilde{x}^\nu}
	+ \omega_{\rho\sigma} \frac{\p x^\rho}{\p \tilde{x}^\mu} \frac{\p x^\sigma}{\p \tilde{x}^\nu},
\end{align}
and the passive transformation law is
\begin{align}
	d_2 x^\mu \rightarrow d_2\tilde{x}^\mu 
	&= 
	\frac{\p \tilde{x}^\mu}{\p x^\rho} d_2 x^\rho 
	+ 
	\frac{\p^2 \tilde{x}^\mu}{\p x^\rho \p x^\sigma} 
	dx^\rho \cdot dx^\sigma,\nonumber\\
	d x^\mu \cdot d x^\nu \rightarrow d\tilde{x}^\mu \cdot d\tilde{x}^\nu 
	&=
	\frac{\p \tilde{x}^\mu}{\p x^\rho} 
	\frac{\p \tilde{x}^\nu}{\p x^\sigma}
	dx^\rho \cdot dx^\sigma. 
\end{align}
The transformation laws should leave the duality pairing \eqref{eq:InProd} invariant. Indeed we find\footnote{One can use the Christoffel symbols to make the second term in the second line vanish.}
\begin{align}
	\langle V, \Omega \rangle 
	&= v^\mu \omega_\mu + v^{\mu\nu}\omega_{\mu\nu} \nonumber\\
	&= \tilde{v}^{\rho} 
	\frac{\p x^\mu}{\p \tilde{x}^\rho} 
	\frac{\p \tilde{x}^\sigma}{\p x^\mu} 
	\tilde{\omega}_{\sigma}
	+ 
	\tilde{v}^{\rho\sigma} 
	\frac{\p^2 x^\mu}{\p \tilde{x}^\rho \p \tilde{x}^\sigma}
	\frac{\p \tilde{x}^\kappa}{\p x^\mu} 
	\tilde{\omega}_{\kappa}
	+ 
	\tilde{v}^{\rho\sigma} 
	\frac{\p x^\mu}{\p \tilde{x}^\rho}
	\frac{\p x^\nu}{\p \tilde{x}^\sigma}
	\frac{\p^2 \tilde{x}^\kappa}{\p x^\mu \p x^\nu} 
	\tilde{\omega}_{\kappa}
	+ 
	\tilde{v}^{\rho\sigma} 
	\frac{\p x^\mu}{\p \tilde{x}^\rho} 
	\frac{\p x^\nu}{\p \tilde{x}^\sigma}
	\frac{\p \tilde{x}^\kappa}{\p x^\mu}
	\frac{\p \tilde{x}^\lambda}{\p x^\nu}
	\tilde{\omega}_{\kappa\lambda} \nonumber\\
	&= \tilde{v}^\mu \tilde{\omega}_\mu + \tilde{v}^{\mu\nu}\tilde{\omega}_{\mu\nu}.
\end{align}

\subsection{Covariance}
In previous subsection, we found that vectors and forms in second order geometry transform in an affine but not contravariant/covariant way. This can be fixed by introducing a covariant basis $\{\hat{\p}_\mu,\hat{\p}_{\mu\nu}\}$ for $\tilde{T}_x\M$ such that
\begin{equation}
	V 
	= \hat{v}^{^\mu_{\nu\rho}} \hat{\p}_{^\mu_{\nu\rho}} 
	= \hat{v}^\mu \hat{\p}_\mu + \hat{v}^{\nu\rho} \, \hat{\p}_{\nu\rho},
\end{equation}
and
\begin{align}
	\hat{\p}_\mu &:= \p_\mu,\nonumber\\
	\hat{\p}_{\mu\nu} &:= \p_{\mu\nu} - \Gamma^{\rho}_{\mu\nu} \p_\rho,\nonumber\\
	\hat{v}^\mu &:= v^\mu + v^{\rho\sigma} \Gamma^{\mu}_{\rho\sigma},\nonumber\\
	\hat{v}^{\mu\nu} &:= v^{\mu\nu}.
\end{align}
In a similar way, we can introduce a contravariant basis for the cotangent space $\tilde{T}^\ast_x\M$, such that
\begin{equation}
	\Omega 
	= \hat{\omega}_{^{\mu}_{\nu\rho}}\, d_2 \hat{x}^{^{\mu}_{\nu\rho}} 
	= \hat{\omega}_\mu\, d_2 \hat{x}^\mu + \hat{\omega}_{\nu\rho} \, d \hat{x}^\nu \cdot d \hat{x}^\rho
\end{equation}
with
\begin{align}
	d_2\hat{x}^\mu &:= d_2 x^\mu + \Gamma^{\mu}_{\nu\rho} dx^\nu \cdot dx^\rho,\nonumber\\
	d \hat{x}^\mu \cdot d \hat{x}^\nu &:= dx^\mu \cdot dx^\nu,\nonumber\\
	\hat{\omega}_\mu &:= \omega_\mu,\nonumber\\
	\hat{\omega}_{\mu\nu} &:= \omega_{\mu\nu} - \omega_{\rho} \Gamma^{\rho}_{\mu\nu}.
\end{align}
\par
It is possible to extend the notion of vector fields and forms to arbitrary $(k,l)$-tensor fields. Indeed, one can construct mappings
\begin{equation}
	T: (\tilde{T}\M)^{\otimes k} \otimes (\tilde{T}^\ast\M)^{\otimes l} \rightarrow \R.
\end{equation}
In local coordinates such a tensor will be given by
\begin{align}\label{eq:TensorRep}
	T 
	&= 
	T^{\left(^\mu_{\nu\rho}\right)_1 ... \left(^\mu_{\nu\rho}\right)_k}
	_{\left(^\sigma_{\kappa\lambda}\right)_1 ... \left(^\sigma_{\kappa\lambda}\right)_l}
	\quad
	\p_{\left(^\mu_{\nu\rho}\right)_1} 
	\otimes ... 
	\otimes \p_{\left(^\mu_{\nu\rho}\right)_k} 
	\otimes d_2 x^{\left(^\sigma_{\kappa\lambda}\right)_1} 
	\otimes ...
	\otimes d_2 x^{\left(^\sigma_{\kappa\lambda}\right)_l}\nonumber\\
	&= 
	T^{\mu_1 ... \mu_k}
	_{\sigma_1 ... \sigma_l} 
	\quad
	\p_{\mu_1} 
	\otimes ... 
	\otimes \p_{\mu_k} 
	\otimes d_2 x^{\sigma_1} 
	\otimes ...
	\otimes d_2 x^{\sigma_l}\nonumber\\
	&\quad +
	T^{(\nu\rho)_1 \mu_2 ... \mu_k}_{\sigma_1 ... \sigma_l}
	\quad
	\p_{\nu_1\rho_1} 
	\otimes \p_{\mu_2} 
	\otimes ... 
	\otimes \p_{\mu_k} 
	\otimes d_2 x^{\sigma_1} 
	\otimes ...
	\otimes d_2 x^{\sigma_l}\nonumber\\
	&\qquad +
	T^{\mu_1 (\nu\rho)_2 \mu_3 ... \mu_k}
	_{\sigma_1 ... \sigma_l}
	\quad
	\p_{\mu_1} 
	\otimes \p_{\nu_2\rho_2} 
	\otimes \p_{\mu_3} 
	\otimes ... 
	\otimes \p_{\mu_k} 
	\otimes d_2 x^{\sigma_1} 
	\otimes ...
	\otimes d_2 x^{\sigma_l}\nonumber\\
	&\qquad + 
	... \nonumber\\
	&\qquad +
	T^{\mu_1 ... \mu_k}
	_{\sigma_1 ... \sigma_{l-1} (\kappa\lambda)_l}
	\quad
	\p_{\mu_1} 
	\otimes ... 
	\otimes \p_{\mu_k} 
	\otimes d_2 x^{\sigma_1} 
	\otimes ...
	\otimes d_2 x^{\sigma_{l-1}}
	\otimes d x^{\kappa_l} \cdot d x^{\lambda_l}\nonumber\\
	&\quad +
	T^{(\nu\rho)_1 (\nu\rho)_2 \mu_3 ... \mu_k}
	_{\sigma_1 ... \sigma_l}
	\quad
	\p_{\nu_1\rho_1} 
	\otimes \p_{\nu_2\rho_2} 
	\otimes \p_{\mu_3} 
	\otimes ... 
	\otimes \p_{\mu_k} 
	\otimes d_2 x^{\sigma_1} 
	\otimes ...
	\otimes d_2 x^{\sigma_l}\nonumber\\
	&\quad +
	...\nonumber\\
	&\quad +
	T^{(\nu\rho)_1 ... (\nu\rho)_k}
	_{(\kappa\lambda)_1 ... (\kappa\lambda)_l}
	\quad
	\p_{\nu_1\rho_1} 
	\otimes ...
	\otimes \p_{\nu_k\rho_k} 
	\otimes d x^{\kappa_1} \cdot d x^{\lambda_1}
	\otimes ...
	\otimes d x^{\kappa_l} \cdot d x^{\lambda_l}.
\end{align}
The components of $T$ do not transform in a covariant/contravariant way. However, one can construct a representation with components $\hat{T}$ such that
\begin{equation}\label{eq:TensorRepCov}
	T 
	= \hat{T}^{\left(^\mu_{\nu\rho}\right)_1 ... \left(^\mu_{\nu\rho}\right)_k}
	_{\left(^\sigma_{\kappa\lambda}\right)_1 ... \left(^\sigma_{\kappa\lambda}\right)_l} 
	\quad
	\hat{\p}_{\left(^\mu_{\nu\rho}\right)_1} 
	\otimes ... 
	\otimes \hat{\p}_{\left(^\mu_{\nu\rho}\right)_k} 
	\otimes d_2 \hat{x}^{\left(^\sigma_{\kappa\lambda}\right)_1} 
	\otimes ...
	\otimes d_2 \hat{x}^{\left(^\sigma_{\kappa\lambda}\right)_l}.
\end{equation}
If expanded as in eq.~\eqref{eq:TensorRepCov}, the coefficients $\hat{T}$ do transform covariantly/contravariantly. The relation between components $T$ and $\hat{T}$ for a general $(k,l)$-tensor can then be derived from the transformation laws for $(1,0)$- and $(0,1)$-tensors.
\par

Finally, we note that there exists a relation between the second order contravariant vectors and covariant forms and the maps $\mathcal{F},\mathcal{G},\mathcal{H}$. For $V\in\tilde{T}_x\M$ we have
\begin{align}
	\mathcal{F}(V) 
	&= \left(v^\mu + v^{\rho\sigma} \Gamma^\mu_{\rho\sigma} \right) \, \p_\mu \nonumber\\
	&= \hat{v}^\mu \, \hat{\p}_\mu, \\
	\mathcal{H}^{\ast}(V) 
	&= v^{\mu\nu} \, \p_\mu \otimes \p_\nu \nonumber\\
	&= \hat{v}^{\mu\nu} \, \hat{\p}_\mu \otimes \hat{\p}_\nu, 
\end{align}
and for $\alpha,\beta\in T^\ast_x\M$
\begin{align}
	\mathcal{G}(\alpha) 
	&= \alpha_\mu \, \left(d_2x^\mu + \Gamma^\mu_{\rho\sigma} dx^\rho \cdot dx^\sigma\right) \nonumber\\
	&= \hat{\alpha}_\mu\, d_2 \hat{x}^\mu, \\
	\mathcal{H}(\alpha\otimes \beta) 
	&= \alpha_\mu \beta_\nu \, dx^\mu \cdot dx^\nu \nonumber\\
	&= \hat{\alpha}_\mu \hat{\beta}_\nu \, d \hat{x}^\mu \cdot d \hat{x}^\nu.
\end{align}
Therefore, all second order vectors and forms can be decomposed into first order vectors, forms and symmetric bilinear tensor products of first order vectors and forms. More generally, any second order $(k,l)$-tensor can be decomposed into first order tensors of degree $(\kappa,\lambda)$ with $k\leq\kappa\leq2k$ and $l\leq\lambda\leq2l$.

\subsection{Second order metric}\label{sec:metric}
In this subsection, we extend the notion of a metric to the second order geometry framework.
We can define a symmetric bilinear function $\tilde{g}:\tilde{T}_x\M\otimes\tilde{T}_x\M\rightarrow \R$, that we call the second order metric tensor. Analogously to the first order metric, it acts on two second order vectors $V,W\in\tilde{T}_x\M$, such that
\begin{align}
	\tilde{g}(V,W) 
	&= \langle V| W\rangle.
\end{align}
Moreover, it induces an isomorphism between vectors and forms
\begin{equation}\label{eq:MetricDefIso}
	\tilde{g}^{\musFlat} : \tilde{T}_x\M \rightarrow \tilde{T}_x^\ast\M
	\qquad {\rm s.t.} \qquad
	\begin{cases}
		\langle V | W \rangle = \langle \tilde{g}^{\musFlat}(V),W \rangle,\\
		\langle \Omega,\Theta \rangle = \langle  \Omega | \tilde{g}^{\musSharp}(\Theta) \rangle.
	\end{cases}
\end{equation}
In a local coordinate chart the metric tensor $\tilde{g}$ can be written as
\begin{align}
	\tilde{g}
	&= 
	\tilde{g}_{\left(^\mu_{\rho\sigma}\right)\left(^\nu_{\kappa\lambda}\right)}
	\; d_2 x^{\left(^\mu_{\rho\sigma}\right)} \otimes d_2 x^{\left(^\nu_{\kappa\lambda}\right)} \nonumber\\
	&= 
	\tilde{g}_{\mu\nu}
	\; d_2 x^\mu \otimes d_2 x^\nu 
	+
	\tilde{g}_{\mu(\kappa\lambda)}
	\; d_2 x^\mu \otimes dx^\kappa \cdot dx^\lambda \nonumber\\
	&\qquad
	+
	\tilde{g}_{(\rho\sigma)\nu}
	\; d x^\rho \cdot dx^\sigma \otimes d_2x^\nu
	+
	\tilde{g}_{(\rho\sigma)(\kappa\lambda)}
	\; d x^\rho \cdot dx^\sigma \otimes dx^\kappa \cdot dx^\lambda.
\end{align}
Using the defining isomorphism \eqref{eq:MetricDefIso} and the duality pairing, eq.~\eqref{eq:InProd}, we find the rules for transforming second order vectors into second order forms:
\begin{align}
	\tilde{g}_{\left(^\mu_{\rho\sigma}\right)\left(^\nu_{\kappa\lambda}\right)} \;
	v^{^\nu_{\kappa\lambda}}
	&=
	v_{^\mu_{\rho\sigma}},\nonumber\\
	\tilde{g}_{\mu\nu}\; v^{\nu}
	+ \tilde{g}_{\mu(\kappa\lambda)}\; v^{\kappa\lambda}
	&= v_\mu, \nonumber\\
	\tilde{g}_{(\rho\sigma)\nu} \; v^{\nu}
	+ \tilde{g}_{(\rho\sigma)(\kappa\lambda)}\; v^{\kappa\lambda} 
	&= v_{\rho\sigma}.
\end{align}
Furthermore, the inverse $\tilde{g}^{-1}$ can be used to transform second order forms into second order vectors:
\begin{align}
	\tilde{g}^{\left(^\mu_{\rho\sigma}\right)\left(^\nu_{\kappa\lambda}\right)} \;
	\omega_{^\nu_{\kappa\lambda}}
	&=
	\omega^{^\mu_{\rho\sigma}},\nonumber\\
	\tilde{g}^{\mu\nu}\; \omega_{\nu}
	+ \tilde{g}^{\mu(\kappa\lambda)}\; \omega_{\kappa\lambda}
	&= \omega^\mu, \nonumber\\
	\tilde{g}^{(\rho\sigma)\nu} \; \omega_{\nu}
	+ \tilde{g}^{(\rho\sigma)(\kappa\lambda)}\; \omega_{\kappa\lambda} 
	&= \omega^{\rho\sigma}.
\end{align}
\par

The components of the metric tensor do not transform covariantly. Therefore, we define a covariant representation of the second order metric:
\begin{align}
	\tilde{g} 
	&= 
	\tilde{g}_{\left(^\mu_{\rho\sigma}\right)\left(^\nu_{\kappa\lambda}\right)}
	\; d_2x^{^\mu_{\rho\sigma}} \otimes d_2x^{^\nu_{\kappa\lambda}} \nonumber\\
	&= 
	\tilde{g}_{\mu\nu}
	\; d_2 \hat{x}^\mu \otimes d_2 \hat{x}^\nu \nonumber\\
	&\quad
	+
	\left(\tilde{g}_{\mu(\kappa\lambda)}
	- \tilde{g}_{\mu\nu} \, \Gamma^\nu_{\kappa\lambda} \right)
	\, d_2 \hat{x}^\mu \otimes d \hat{x}^\kappa \cdot d \hat{x}^\lambda \nonumber\\
	&\quad
	+
	\left( \tilde{g}_{(\rho\sigma)\nu}
	- \tilde{g}_{\mu\nu} \, \Gamma^\mu_{\rho\sigma} \right)
	\, d \hat{x}^\rho \cdot d \hat{x}^\sigma \otimes d_2 \hat{x}^\nu \nonumber\\
	&\quad
	+
	\left(\tilde{g}_{(\rho\sigma)(\kappa\lambda)}
	+ \tilde{g}_{\mu\nu} \, \Gamma^\mu_{\rho\sigma}\, \Gamma^\nu_{\kappa\lambda}
	- \tilde{g}_{\mu(\kappa\lambda)} \, \Gamma^\mu_{\rho\sigma}
	- \tilde{g}_{(\rho\sigma)\nu} \, \Gamma^\nu_{\kappa\lambda} \right)
	\, d \hat{x}^\rho \cdot d \hat{x}^\sigma \otimes d \hat{x}^\kappa \cdot d \hat{x}^\lambda\nonumber\\
	&= 
	\hat{g}_{\mu\nu}
	\; d_2 \hat{x}^\mu \otimes d_2 \hat{x}^\nu 
	+
	\hat{g}_{\mu(\kappa\lambda)}
	\; d_2 \hat{x}^\mu \otimes d \hat{x}^\kappa \cdot d \hat{x}^\lambda \nonumber\\
	&\qquad
	+
	\hat{g}_{(\rho\sigma)\nu}
	\; d \hat{x}^\rho \cdot d \hat{x}^\sigma \otimes d_2 \hat{x}^\nu
	+
	\hat{g}_{(\rho\sigma)(\kappa\lambda)}
	\; d \hat{x}^\rho \cdot d \hat{x}^\sigma \otimes d \hat{x}^\kappa \cdot d \hat{x}^\lambda\nonumber\\
	&= 
	\hat{g}_{\left(^\mu_{\rho\sigma}\right)\left(^\nu_{\kappa\lambda}\right)}
	\; d_2 \hat{x}^{^\mu_{\rho\sigma}} \otimes d_2 \hat{x}^{^\nu_{\kappa\lambda}}.
\end{align}
We notice that a second order vector can be uniquely decomposed in a first order vector and a bilinear first order tensor. We will therefore impose
\begin{equation}
	\tilde{g}^{\musFlat} = 
	\begin{pmatrix}
		\mathcal{G} \circ g^{\musFlat} \circ \mathcal{F} \\
		\mathcal{H} \circ \left( g^{\musFlat} \otimes g^{\musFlat}\right) \circ \mathcal{H}^\ast
	\end{pmatrix}
\end{equation}
We can then write in a local coordinate system
\begin{align}\label{eq:CovComponentsSecMetric}
	\hat{g}_{\left(^\mu_{\rho\sigma}\right)\left(^\nu_{\kappa\lambda}\right)}
	&=
	\begin{pmatrix}
		\hat{g}_{\mu\nu} & \hat{g}_{\mu(\kappa\lambda)} \\
		\hat{g}_{(\rho\sigma)\nu} & \hat{g}_{(\rho\sigma)(\kappa\lambda)}
	\end{pmatrix}
	\nonumber\\
	&=
	\begin{pmatrix}
		g_{\mu\nu} 
		& 0 \\
		0 
		& \frac{1}{2}\left( g_{\rho\kappa} g_{\sigma\lambda} + g_{\rho\lambda} g_{\sigma\kappa} \right)
	\end{pmatrix}
\end{align}
where we have suppressed the maps $\mathcal{F}$, $\mathcal{G}$, $\mathcal{H}$, $\mathcal{H}^\ast$ in the second line and where $g_{\mu\nu}$ are the components of the first order metric. The inverse can be written as
\begin{align}\label{eq:MetricCovSecond}
	\hat{g}^{\left(^\mu_{\rho\sigma}\right)\left(^\nu_{\kappa\lambda}\right)}
	&=
	\begin{pmatrix}
		\hat{g}^{\mu\nu} & \hat{g}^{\mu(\kappa\lambda)} \\
		\hat{g}^{(\rho\sigma)\nu} & \hat{g}^{(\rho\sigma)(\kappa\lambda)}
	\end{pmatrix}
	\nonumber\\
	&=
	\begin{pmatrix}
		g^{\mu\nu} 
		& 0 \\
		0 
		& \frac{1}{2}\left( g^{\rho\kappa} g^{\sigma\lambda} + g^{\rho\lambda} g^{\sigma\kappa} \right)
	\end{pmatrix}
\end{align}
We can now raise and lower indices on covariant forms and contravariant vectors in the usual way
\begin{align}
	\hat{g}_{\left(^\mu_{\rho\sigma}\right)\left(^\nu_{\kappa\lambda}\right)} \;
	\hat{v}^{^\nu_{\kappa\lambda}}
	&=
	\hat{v}_{^\mu_{\rho\sigma}},\nonumber\\
	\hat{g}_{\mu\nu} \hat{v}^{\nu}
	&= \hat{v}_\mu, \nonumber\\
	\hat{g}_{(\rho\sigma)(\kappa\lambda)} \hat{v}^{\kappa\lambda} 
	&= \hat{v}_{\rho\sigma},\nonumber\\
	\hat{g}^{\left(^\mu_{\rho\sigma}\right)\left(^\nu_{\kappa\lambda}\right)} \;
	\hat{\omega}_{^\nu_{\kappa\lambda}}
	&=
	\hat{\omega}^{^\mu_{\rho\sigma}},\nonumber\\
	\hat{g}^{\mu\nu} \hat{\omega}_{\nu}
	&= \hat{\omega}^\mu, \nonumber\\
	\hat{g}^{(\rho\sigma)(\kappa\lambda)} \hat{\omega}_{\kappa\lambda}
	&= \hat{\omega}^{\rho\sigma},
\end{align}
where we used the symmetry of $v^{\mu\nu}$ and $\omega_{\mu\nu}$. Finally we can express the second order metric components $\tilde{g}$ in terms of the first order metric:
\begin{align}
	\tilde{g}_{\left(^\mu_{\rho\sigma}\right)\left(^\nu_{\kappa\lambda}\right)}
	&=
	\begin{pmatrix}
		\tilde{g}_{\mu\nu} & \tilde{g}_{\mu(\kappa\lambda)} \\
		\tilde{g}_{(\rho\sigma)\nu} & \tilde{g}_{(\rho\sigma)(\kappa\lambda)}
	\end{pmatrix}
	\nonumber\\
	&=
	\begin{pmatrix}
		g_{\mu\nu} 
		& g_{\mu\alpha}\,\Gamma^{\alpha}_{\kappa\lambda} \\
		g_{\alpha\nu}\,\Gamma^{\alpha}_{\rho\sigma}
		& \frac{1}{2}\left( g_{\rho\kappa} g_{\sigma\lambda} + g_{\rho\lambda} g_{\sigma\kappa} \right)
		+ g_{\alpha\beta}\,\Gamma^{\alpha}_{\rho\sigma}\,\Gamma^{\beta}_{\kappa\lambda}
	\end{pmatrix}
\end{align}
Its inverse is given by
\begin{align}
	\tilde{g}^{\left(^\mu_{\rho\sigma}\right)\left(^\nu_{\kappa\lambda}\right)}
	&=
	\begin{pmatrix}
		\tilde{g}^{\mu\nu} & \tilde{g}^{\mu(\kappa\lambda)} \\
		\tilde{g}^{(\rho\sigma)\nu} & \tilde{g}^{(\rho\sigma)(\kappa\lambda)}
	\end{pmatrix}
	\nonumber\\
	&=
	\begin{pmatrix}
		g^{\mu\nu} 
		+ g^{\alpha\eta} g^{\beta\xi} \Gamma^\mu_{\alpha\beta} \Gamma^\nu_{\eta\xi}
		& 
		- g^{\alpha\kappa} g^{\beta\lambda} \Gamma^\mu_{\alpha\beta} \\
		- g^{\rho\alpha} g^{\sigma\beta} \Gamma^\nu_{\alpha\beta}
		& \frac{1}{2}\left( g^{\rho\kappa} g^{\sigma\lambda} + g^{\rho\lambda} g^{\sigma\kappa} \right)
	\end{pmatrix}
\end{align}

\subsection{$k$-forms}
In this subsection, we extend the notion of $k$-forms to the second order geometry framework.
As usual, we denote the bundle of covariant $k$-tensors by $T^k(T^\ast\M)$ and the subbundle of alternating $k$-tensors by $\Lambda^k(T^\ast\M)$. The rank of the latter bundle is $n\choose{k}$ and a $k$-form $\omega\in\Lambda^k(T^\ast\M)$ can be written as
\begin{equation}
	\omega = \omega_{\mu_1...\mu_k}\, dx^{\mu_1} \wedge ... \wedge dx^{\mu_k}
\end{equation}
where we assume $\mu_1<...<\mu_k$. Similarly, we construct a bundle of second order $k$-tensors $T^k(\tilde{T}^\ast\M)$ and a subbundle $\Lambda^k(\tilde{T}^\ast\M)$ of rank $N\choose{k}$ with $N=\frac{1}{2}n(n+3)$. A  second order $k$-form $\Omega\in\Lambda^k(T^\ast\M)$ can be written as
\begin{align}
	\Omega 
	&= \omega_{\left(^{\mu}_{\nu\rho}\right)_1...\left(^{\mu}_{\nu\rho}\right)_k}
	\, d_2x^{\left(^{\mu}_{\nu\rho}\right)_1} \wedge ... \wedge d_2x^{\left(^{\mu}_{\nu\rho}\right)_k}\nonumber\\
	&= 
	\omega_{\mu_1...\mu_k}
	\, d_2 x^{\mu_1} \wedge ... \wedge d_2 x^{\mu_k}\nonumber\\
	&\quad 
	+ \omega_{(\nu\rho)_1\mu_2...\mu_k}
	\, d x^{\nu_1} \cdot d x^{\rho_1}  \wedge d_2 x^{\mu_2} \wedge ... \wedge d_2 x^{\mu_k}\nonumber\\
	&\qquad
	+ \omega_{\mu_1(\nu\rho)_2\mu_3...\mu_k}
	\, d_2 x^{\mu_1} \wedge d x^{\nu_2} \cdot d x^{\rho_2}  \wedge d_2x^{\mu_3} \wedge ... \wedge d_2 x^{\mu_k}\nonumber\\
	&\qquad
	+ ... \nonumber\\
	&\qquad
	+ \omega_{\mu_1\mu_2...\mu_{k-1}(\nu\rho)_k}
	\, d_2 x^{\mu_1} \wedge  d_2 x^{\mu_2} \wedge ... \wedge d_2 x^{\mu_{k-1}} \wedge d x^{\nu_k} \cdot d x^{\rho_k}\nonumber\\
	&\quad
	+ \omega_{(\nu\rho)_1(\nu\rho)_2\mu_3...\mu_k}
	\, d x^{\nu_1} \cdot d x^{\rho_1} \wedge d x^{\nu_2} \cdot d x^{\rho_2} \wedge d_2 x^{\mu_3} \wedge  ... \wedge d_2 x^{\mu_k} \nonumber\\
	& \quad
	+ ... \nonumber\\
	& \quad
	+ \omega_{(\nu\rho)_1...(\nu\rho)_k}
	\, d x^{\nu_1} \cdot d x^{\rho_1} \wedge  ... \wedge d x^{\nu_k} \cdot d x^{\rho_k}.
\end{align}

\subsection{Exterior derivatives}
In this subsection, we extend the notion of the exterior derivative to the second order geometry framework.
The first order exterior derivative is a map $d: \Lambda^k(T^\ast\M) \rightarrow \Lambda^{k+1}(T^\ast\M)$ such that
\begin{equation}
	d \omega = \p_{\nu} \omega_{\mu_1...\mu_k} \, dx^\nu \wedge dx^{\mu_1} \wedge ... \wedge dx^{\mu_k},
\end{equation}
which is linear:
\begin{align}
	d(\omega + \theta) &= d\omega + d\theta \qquad \forall\; \omega,\theta \in \Lambda^k(T^\ast\M),\nonumber\\
	d(c\,\omega) &= c\; d\omega \qquad \forall\; \omega \in \Lambda^k(T^\ast\M),\, c\in\R; 
\end{align}
satisfies the modified Leibniz rule:
\begin{equation}
	d(\omega \wedge \theta) = d\omega \wedge \theta + (-1)^k \omega \wedge d\theta \qquad \forall\; \omega \in \Lambda^k(T^\ast\M),\, \theta \in \Lambda^l(T^\ast\M);
\end{equation}
satisfies the closure condition
\begin{equation}
	d(d(\omega)) = 0 \qquad \forall\,\omega \in \Lambda^k(T^\ast\M);
\end{equation}
and commutes with pullbacks:
\begin{equation}
	\phi^\ast(d\omega) = d(\phi^\ast(\omega)) \qquad \forall\; \omega\in \Lambda^k(T^\ast\M),\, \phi\in C^{\infty}(\M,\R).
\end{equation}
Analagously we define a a second order exterior derivative $d_2:\Lambda^k(\tilde{T}^\ast\M) \rightarrow \Lambda^{k+1}(\tilde{T}^\ast\M)$ such that
\begin{align}
	d_2\, \Omega &= \p_{^\nu_{\kappa\lambda}}\, \omega_{\left(^{\mu}_{\rho\sigma}\right)_1...\left(^{\mu}_{\rho\sigma}\right)_k} \, d_2x^{^{\nu}_{\kappa\lambda}} \wedge d_2x^{\left(^{\mu}_{\rho\sigma}\right)_1} \wedge ... \wedge d_2x^{\left(^{\mu}_{\rho\sigma}\right)_k} \nonumber\\
	&= \p_\nu \, \omega_{\left(^{\mu}_{\rho\sigma}\right)_1...\left(^{\mu}_{\rho\sigma}\right)_k}
	\, d_2x^\nu \wedge d_2x^{\left(^{\mu}_{\rho\sigma}\right)_1} \wedge ... \wedge d_2x^{\left(^{\mu}_{\rho\sigma}\right)_k} \nonumber\\
	&\quad
	+ \p_\kappa \p_\lambda\, \omega_{\left(^{\mu}_{\rho\sigma}\right)_1...\left(^{\mu}_{\rho\sigma}\right)_k}
	\, dx^\kappa \cdot dx^\lambda \wedge d_2x^{\left(^{\mu}_{\rho\sigma}\right)_1} \wedge ... \wedge d_2x^{\left(^{\mu}_{\rho\sigma}\right)_k}.
\end{align}
This second order exterior derivative is also linear and commutes withs pullbacks. Furthermore, it obeys the closure condition
\begin{equation}
	d_2(d_2(\Omega)) = 0 \qquad \forall\,\Omega \in \Lambda^k(\tilde{T}^\ast\M);
\end{equation}
and a new modified Leibniz rule
\begin{equation}
	d_2(\Omega \wedge \Theta) 
	= d_2\Omega \wedge \Theta 
	+ (-1)^k \Omega \wedge d_2\Theta
	+ 2 d\Omega \cdot d\Theta
	\qquad \forall\; \Omega \in \Lambda^k(\tilde{T}^\ast\M),\, \Theta \in \Lambda^l(\tilde{T}^\ast\M),
\end{equation}
where
\begin{equation}
	d\Omega \cdot d\Theta
	= \p_\alpha\, \omega_{\left(^{\mu}_{\rho\sigma}\right)_1...\left(^{\mu}_{\rho\sigma}\right)_k}
	\, \p_\beta\, \omega_{\left(^{\nu}_{\kappa\lambda}\right)_1...\left(^{\nu}_{\kappa\lambda}\right)_l}
	\, dx^\alpha \cdot dx^\beta \wedge 
	d_2x^{\left(^{\mu}_{\rho\sigma}\right)_1} \wedge ... \wedge d_2x^{\left(^{\mu}_{\rho\sigma}\right)_k}
	\wedge d_2x^{\left(^{\nu}_{\kappa\lambda}\right)_1} \wedge ... \wedge d_2x^{\left(^{\nu}_{\kappa\lambda}\right)_l}.
\end{equation}
The proof for these properties is similar to the proof for the corresponding properties in first order geometry, and is therefore omitted.

\subsection{Interior products}
In this subsection, we extend the notion of the interior product to the second order geometry framework.
The first order interior product is a map $\iota_v:\Lambda^k(T^\ast\M)\rightarrow \Lambda^{k-1}(T^\ast\M)$ such that
\begin{equation}
	\iota_v \, \omega 
	= 
	\sum_{l=1}^k (-1)^{l-1}\, v^{\mu_l}\, \omega_{\mu_1...\mu_k} \, dx^{\mu_1} \wedge ... \wedge dx^{\mu_{l-1}} \wedge dx^{\mu_{l+1}} \wedge ... \wedge dx^{\mu_k}.
\end{equation}
This map is linear, commutes with pullbacks, satisfies the modified Leibniz rule and satisfies the anti-symmetry property
\begin{equation}
	\{ \iota_u, \iota_v\}\, \omega = 0.
\end{equation}
Similarly, one can define a second order interior product $\iota_V:\Lambda^k(T^\ast\M)\rightarrow \Lambda^{k-1}(T^\ast\M)$, such that
\begin{equation}
	\iota_V\, \Omega 
	= \sum_{l=1}^k (-1)^{l-1}\, v^{\left(^\mu_{\rho\sigma}\right)_l}\; \omega_{\left(^\mu_{\rho\sigma}\right)_1...\left(^\mu_{\rho\sigma}\right)_k} \, d_2x^{\left(^\mu_{\rho\sigma}\right)_1} \wedge ... \wedge d_2x^{\left(^\mu_{\rho\sigma}\right)_{l-1}} \wedge d_2x^{\left(^\mu_{\rho\sigma}\right)_{l+1}} \wedge ... \wedge d_2x^{\left(^\mu_{\rho\sigma}\right)_k},
\end{equation}
which satisfies the same properties with the modified Leibniz rule replaced by a new modified Leibniz rule as in previous subsection.

\subsection{Lie derivatives}
Using the results from previous subsections, we can extend the notion of a Lie derivative to the second order geometry framework.
A family of diffeomorphisms $\phi_\lambda := \R \times \M \rightarrow \M$ satisfying the usual (semi-)group properties can be thought of as a vector field $v\in \mathfrak{X}(\M)$ that generates a set of integral curves $\gamma_v:\R\rightarrow\M$ along the vector field. Along any such integral curve parametrized by $\lambda$, one can define the first order derivative of a function $f\in C^\infty(\M,\R)$ by
\begin{equation}\label{eq:VecDer}
	\frac{d}{d\lambda} f = \frac{d x^\mu}{d\lambda} \p_\mu f = v^\mu \p_\mu f = v\,f.
\end{equation}
This derivative is equivalent to the Lie derivative along the vector field $v$
\begin{equation}\label{LieScal}
	\Ls_vf = v\, f,
\end{equation} 
which can be generalized to a Lie derivative acting on vectors and forms given by
\begin{align}
	\Ls_{v} u &= [v,u],\nonumber\\
	\Ls_{v} \omega &= \{\iota_v, d\} \omega.
\end{align}
In a local coordinate chart, these expressions can be written as
\begin{align}
	\Ls_{v} u^\mu &= v^\nu \p_\nu u^\mu - u^\nu \p_\nu v^\mu,\\
	\Ls_{v} \omega_\mu &= v^\nu \p_\nu \omega_\mu + (\p_\mu v^\nu)\omega_\nu.
\end{align}
Furthermore, using the Leibniz rule one can construct Lie derivatives acting on arbitrary tensor fields.
\par

We can analogously define a notion of a Lie derivative of second order tensors along a second order vector field $V\in\tilde{\mathfrak{X}}(\M)$. As defining relations for derivatives of vectors $U\in\tilde{T}_x\M$ and forms $\Omega\in\tilde{T}^\ast_x\M$ we take
\begin{align}
	\Ls_V f &= V f,\nonumber\\
	\Ls_{V} U &= [V,U],\nonumber\\
	\Ls_{V} \Omega &= \{\iota_V, d_2\} \Omega.
\end{align}
In order to make these expressions well defined, we impose
\begin{align}
	v^{\mu\sigma}\p_\sigma u^{\nu\rho} 
	&= u^{\mu\sigma}\p_\sigma v^{\nu\rho},\\
	\omega_{\mu\nu}
	&= \p_\mu\omega_\nu.
\end{align}
In order to satisfy the first condition, we impose $u^{\mu\nu}=k\, v^{\mu\nu}$ with $k\in\R$ and define $W\in T\M\subset \tilde{T}\M$ such that
\begin{equation}
	W = k V - U = 
	\begin{pmatrix}
		k v^\mu - u^\mu\\
		0
	\end{pmatrix}
\end{equation}
In local coordinates we then find
\begin{align}
	\Ls_{V} f 
	&= \left(v^\sigma \p_\sigma + v^{\sigma\kappa} \p_\sigma \p_\kappa \right) f,\nonumber\\
	\Ls_{V}\, U^\mu
	&= \left(v^\sigma \p_\sigma + v^{\sigma\kappa} \p_\sigma \p_\kappa \right) u^\mu - u^\sigma \p_\sigma v^\mu - u^{\sigma\kappa} \p_\sigma \p_\kappa v^\mu,\nonumber\\
	\Ls_{V}\, U^{\nu\rho}
	&=  w^\sigma \p_\sigma v^{\nu\rho} - v^{\nu\sigma} \p_\sigma w^\rho - v^{\rho\sigma} \p_\sigma w^\nu,
	\nonumber\\
	\Ls_{V}\, \Omega_\mu
	&= \left( v^\sigma \p_\sigma + v^{\sigma\kappa} \p_\sigma \p_\kappa\right) \omega_\mu + \omega_\sigma \p_\mu v^\sigma + \omega_{\sigma\kappa} \p_\mu v^{\sigma\kappa},\nonumber\\
	\Ls_{V}\, \Omega_{\nu\rho}
	&=	\left( v^\sigma \p_\sigma + v^{\sigma\kappa} \p_\sigma \p_\kappa\right) \omega_{\nu\rho}  + \omega_\sigma \p_\nu \p_\rho v^\sigma + \omega_{\sigma\kappa} \p_\nu \p_\rho v^{\sigma\kappa}\nonumber\\
	&  \quad
	+ 2\p_{(\nu} v^\sigma \p_{\rho)} \omega_\sigma + 2\p_{(\nu} v^{\sigma\kappa} \p_{\rho)} \omega_{\sigma\kappa}.
\end{align}
or equivalently with respect to the covariant bases
\begin{align}
	\Ls_{V} f 
	&= \left(\hat{v}^\sigma \nabla_\sigma + \hat{v}^{\sigma\kappa} \nabla_\sigma \nabla_\kappa \right) f,\nonumber\\
	\Ls_{V}\, \hat{U}^\mu
	&= \left(\hat{v}^\sigma \nabla_\sigma + \hat{v}^{\sigma\kappa} \nabla_\sigma \nabla_\kappa \right) \hat{u}^\mu 
	- \hat{u}^\sigma \nabla_\sigma \hat{v}^\mu 
	- \hat{u}^{\sigma\kappa} \nabla_\sigma \nabla_\kappa \hat{v}^\mu 
	+ \mathcal{R}^\mu_{\;\;\sigma\kappa\lambda} \hat{v}^{\sigma\kappa} \hat{w}^\lambda,\nonumber\\
	\Ls_{V}\, \hat{U}^{\nu\rho}
	&= \hat{w}^\sigma \nabla_\sigma \hat{v}^{\nu\rho} 
	- \hat{v}^{\nu\sigma} \nabla_\sigma \hat{w}^\rho 
	- \hat{v}^{\rho\sigma} \nabla_\sigma \hat{w}^\nu,
	\nonumber\\
	\Ls_{V}\, \hat{\Omega}_\mu
	&= \left( \hat{v}^\sigma \nabla_\sigma + \hat{v}^{\sigma\kappa} \nabla_\sigma \nabla_\kappa\right) \hat{\omega}_\mu + \hat{\omega}_\sigma \nabla_\mu \hat{v}^\sigma 
	+ \hat{\omega}_{\sigma\kappa} \nabla_\mu \hat{v}^{\sigma\kappa} 
	+ \mathcal{R}^\sigma_{\;\;\kappa\lambda\mu} \hat{v}^{\kappa\lambda} \hat{\omega}_{\sigma},\nonumber\\
	\Ls_{V}\, \hat{\Omega}_{\nu\rho}
	&= \left( \hat{v}^\sigma \nabla_\sigma 
	+ \hat{v}^{\sigma\kappa} \nabla_\sigma \nabla_\kappa\right) \hat{\omega}_{\nu\rho} 
	+ \hat{\omega}_\sigma \nabla_{(\nu} \nabla_{\rho)} \hat{v}^\sigma 
	+ \hat{\omega}_{\sigma\kappa} \nabla_{(\nu} \nabla_{\rho)} \hat{v}^{\sigma\kappa}
	+ 2 \nabla_{(\nu|} \hat{v}^\sigma \nabla_{|\rho)} \hat{\omega}_\sigma \nonumber\\
	& \quad  
	+ 2 \nabla_{(\nu|} \hat{v}^{\sigma\kappa} \nabla_{|\rho)} \hat{\omega}_{\sigma\kappa} 
	- \mathcal{R}^\kappa_{\;\;(\nu\rho)\sigma} \hat{v}^\sigma \hat{\omega}_\kappa
	+ 2 \hat{v}^{\sigma\kappa}\left( 
	\mathcal{R}^\lambda_{\;\;\sigma\kappa(\nu} \hat{\omega}_{\rho)\lambda}
	- \mathcal{R}^\lambda_{\;\;(\nu\rho)\sigma} \hat{\omega}_{\kappa\lambda} \right)\nonumber\\
	&\quad - \hat{v}^{\sigma\kappa} \hat{\omega}_\lambda \left( 
	\nabla_\sigma \mathcal{R}^\lambda_{\;\;(\nu\rho)\kappa}
	+ \nabla_{(\nu|} \mathcal{R}^\lambda_{\;\;\sigma|\rho)\kappa} \right).
\end{align}
The Lie derivatives for first order vectors and forms and along first order vector fields can easily be obtained from these formulae by taking the appropriate limit. Only the Lie derivative of a second order vector field along a first order vector field cannot be derived as a limit from these formulae. This one can be obtained by replacing $v^{\mu\nu}\rightarrow u^{\mu\nu}$ and $w^\mu\rightarrow v^\mu$ in the above formulae.

\subsection{Parallel transport}\label{sec:ParTrans}
In this subsection, we discuss the notion of parallel transport along second order vector fields. This notion is similar to the notion of stochastic parallel transport along semi-martingales as developed by Dohrn and Guerra \cite{DohrnGuerraI,DohrnGuerraII}. It is different from first order parallel transport, as the second order part of the vector fields generate geodesic deviation. Here, we closely follow the presentation of stochastic parallel transport by Nelson, cf. section 10 in Ref.~\cite{Nelson}.
\par

Let $X(\tau)$ be a path in $\M$, passing through the points $x,y\in\M$ at times $\tau_1,\tau_2$. We will assume that there exists a convex coordinate chart $(U,\chi)$ such that $x,y\in U$. Moreover, let $V\in\tilde{T}_x\M$ be a second order tangent vector at $x$ with $\hat{v}=\mathcal{F}(V)$ its contravariant first order projection, such that in $\chi(U)$ we have $y^\mu = x^\mu + \hat{v}^\mu$.
\par

Let $d_2\hat{X}(\tau)\in\mathcal{F}(T\M)$ be a transport and let $d_2\hat{x}^\mu=d_2\hat{X}(\tau_1)$ and $d_2\hat{y}^\mu=d_2\hat{X}(\tau_2)$ be its values when passing through $x$ and $y$ respectively. Then, using the standard notion of parallel transport, $d_2\hat{X}(\tau)$ is said to be a parallel transport, if
\begin{equation}\label{eq:ParTrans1}
	d_2 \hat{y}^\mu = d_2 \hat{x}^\mu - \Gamma^\mu_{\rho\sigma}(x)\, \hat{v}^\rho\, d_2 \hat{x}^\sigma.
\end{equation}
In order to extend this notion to second order vector fields, we define the difference vector
\begin{equation}
	d_2\hat{v}^\mu := d_2y^\mu - d_2x^\mu.
\end{equation}
Using the parallel transport equation \eqref{eq:ParTrans1}, the relations
\begin{align}
	d_2\hat{x}^\mu &= d_2x^\mu + \Gamma^\mu_{\rho\sigma}(x)\, d\hat{x}^\rho \cdot d\hat{x}^\sigma,\nonumber\\
	d_2\hat{y}^\mu &= d_2y^\mu + \Gamma^\mu_{\rho\sigma}(y)\, d\hat{y}^\rho \cdot d\hat{y}^\sigma
\end{align}
and the Taylor expansion
\begin{equation}
	\Gamma^\mu_{\rho\sigma}(y) = \Gamma^\mu_{\rho\sigma}(x) + \p_\nu \Gamma^\mu_{\rho\sigma}(x) \hat{v}^\nu + \mathcal{O}(\hat{v}^2),
\end{equation}
we find
\begin{align}
	d_2 \hat{v}^\mu 
	&= 
	- \Gamma^\mu_{\rho\sigma} \hat{v}^\rho d_2x^\sigma
	- \left(
	\p_\nu \Gamma^\mu_{\rho\sigma}
	+ \Gamma^\mu_{\nu\kappa} \Gamma^\kappa_{\rho\sigma}
	- 2 \Gamma^\mu_{\rho\kappa} \Gamma^\kappa_{\nu\sigma} 
	\right) \hat{v}^\nu\, dx^\rho \cdot dx^\sigma\nonumber\\
	&=
	- \Gamma^\mu_{\rho\sigma} \hat{v}^\rho d_2\hat{x}^\sigma
	- \left(
	\p_\nu \Gamma^\mu_{\rho\sigma}
	- 2 \Gamma^\mu_{\rho\kappa} \Gamma^\kappa_{\nu\sigma} 
	\right) \hat{v}^\nu\, d\hat{x}^\rho \cdot d\hat{x}^\sigma
\end{align}
where $\Gamma^\mu_{\rho\sigma}=\Gamma^\mu_{\rho\sigma}(x)$. We will call this the \textit{equation of second order parallel transport}. Notice that the equation of first order parallel transport is obtained if $d\hat{X}\in T\M$ is a first order transport and $V\in T\M$ is a first order vector, as this implies  $dx^\rho \cdot dx^\sigma=0$ and $\hat{v}=v$ respectively.
\par 

The equation of second order parallel transport is linear in $\hat{v}$ and has a solution of the form
\begin{equation}
	\hat{v}^\mu(\tau_2) = P^\mu_{\;\;\nu} (\tau_2,\tau_1)\, \hat{v}^\nu(\tau_1),
\end{equation}
where $P^\mu_{\;\;\nu} (\tau_2,\tau_1)$ is the \textit{second order parallel propagator}. Using this propagator, we can define the \textit{second order directional covariant derivative} $\hat{d}$ by
\begin{align}\label{eq:DirStochCovDev}
	\hat{d}_2 \hat{v}^\mu 
	&=  
	P^\mu_{\;\;\nu} (\tau_1,\tau_2)\, \hat{v}^\nu(\tau_2) - \hat{v}^\mu(\tau_1)\nonumber\\
	&=
	d_2 \hat{v}^\mu
	+ \Gamma^\mu_{\rho\sigma} \hat{v}^\rho d_2\hat{x}^\sigma
	+ \left(
	\p_\nu \Gamma^\mu_{\rho\sigma}
	- 2 \Gamma^\mu_{\rho\kappa} \Gamma^\kappa_{\nu\sigma} 
	\right) \hat{v}^\nu\, d\hat{x}^\rho \cdot d\hat{x}^\sigma.
\end{align}
\subsection{Embeddings into higher dimensions}
As an aside, we discuss the relation between second order geometry and first order geometry on higher dimensional manifolds. One can embed a $n$-dimesional pseudo-Riemannian manifold with signature\footnote{We denote the signature by (+,-,0). i.e. $(d,1,0)$ corresponds to a $(-+...+)$ metric.} $(d,1,0)$ into a $N$-dimensional pseudo-Riemannian manifold $\tilde{M}$ with signature\footnote{More generally, if $\M$ has signature $(k,l,m)$, then $\tilde{\M}$ has signature $(K,L,M)$ with $K=\frac{1}{2}\left[k(k+3)+l(l+1)\right]$, $L=l(k+1)$ and $M=\frac{m}{2}\left(2k+2l+m+3\right)$.} $(D,n,0)$ with $N=\frac{1}{2}n(n+3)$ and $D=\frac{1}{2}n(n+1)$. We can for example take the trivial embedding
\begin{equation}
	\iota:\M \hookrightarrow \tilde{\M} 
	\qquad {\rm s.t.} \qquad
	\begin{cases}
		\iota^\alpha(x) = x^\alpha, & {\rm if}\; \alpha\leq d; \\
		\iota^\alpha(x) = 0, & {\rm if}\; \alpha > d.
	\end{cases}
\end{equation}
\par

The pushforward $\iota_\ast$ of this embedding defines for every $x\in\M$ a bijection between the second order tangent space $\tilde{T}_x\M$ and the first order tangent space $T_{\iota(x)}\tilde{\M}$. Additionally, the pullback $\iota^\ast$ defines a bijection between the cotangent spaces $\tilde{T}_x^\ast\M$ and $T_{\iota(x)}^\ast\tilde{\M}$. This bijection $\iota^\ast$ acts on the basis vectors as\footnote{Notice that $\mu\in{0,1,...,d}$, and that $\rho\leq\sigma$.}
\begin{align}
	d_2 x^\mu &\mapsto dx^\mu, \nonumber\\
	d x^\rho \cdot dx^\sigma &\mapsto dx^{n+\frac{1}{2}\rho(2n-\rho-1)+\sigma}.
\end{align}
Moreover, this induces a bijection between the second order metric on $\M$ and the first order metric on $\tilde{\M}$:
\begin{equation}
	\tilde{g}_{\left(^\mu_{\rho\sigma}\right)\left(^\nu_{\kappa\lambda}\right)} \mapsto 				\tilde{g}_{\alpha\beta}
\end{equation}
with $\alpha,\beta\in\{0,1,...,N\}$. One can thus describe the second order geometry framework using the first order formalism on a $N$-dimensional manifold $\tilde{\M}$ instead of the original $n$-dimensional manifold $\M$. However, the support of functions defined on $\tilde{\M}$ must be restricted to the subspace $\M\subset\tilde{\M}$.

\section{Manifold Valued Semi-Martingales}\label{sec:StochProc}

In this section, we discuss stochastic motion on a manifold. Classically, a particle follows a trajectory or path on the manifold, that is parametrized by its proper time. In other words a trajectory is a map $\gamma: T \rightarrow \M$, where $T=[\tau_i,\tau_f]\subset\R$.
\par

We make this notion stochastic by promoting the manifold to a measurable space $(\M,\mathcal{B}(\M))$, where $\mathcal{B}(\M)$ is the Borel sigma algebra of $\M$. Furthermore, we introduce the probability space $(\Omega,\Sigma,\mathbb{P})$, and the random variable $X:(\Omega,\Sigma,\mathbb{P})\rightarrow(\M,\mathcal{B}(\M))$. Given $T=[\tau_i,\tau_f]\subset\R$ we can introduce a filtration $\{\mathcal{P}_\tau\}_{\tau\in T}$, which is by definition an ordered set such that $\mathcal{P}_{\tau_i}\subseteq\mathcal{P}_s\subseteq\mathcal{P}_t\subseteq\Sigma$ $\forall\,s<t\in T$. In addition, we assume the filtration to be right-continuous, i.e. $\mathcal{P}_\tau=\cap_{\epsilon>0} \mathcal{P}_{\tau+\epsilon}$.
\par 

We can then introduce a stochastic process adapted to this filtration as a family of random variables $\{X(\tau):\tau\in T\}$. We will restrict the set of stochastic processes to the continuous manifold valued semi-martingales. These are the continuous manifold valued stochastic processes $\{X(\tau)\}_{\tau\in T}$ such that $f(X)$ is a semi-martingle for every smooth function $f\in C^\infty(\M,\R^{n})$. In particular, for a coordinate chart $\chi:U\rightarrow V$ with $U\subset\M$ and $V\subset\R^{d+1}$ the coordinates $X^\mu=\chi^\mu(X)$ are semi-martingales. A semi-martingale is a process $X(\tau)$ that can be decomposed as
\begin{equation}
	X(\tau) = x_i + C_+(\tau) + W_+(\tau),
\end{equation}
where $x_i:=X(\tau_i)$ is the initial value of the process, $C_+(\tau)$ is a local c\`adl\`ag process with finite variation, such that $C_+(\tau_i)=0$, and $W_+(\tau)$ is a local martingale process, such that $W_+(\tau_i)=0$, satisfying the martingale property
\begin{equation}
	\E_{t^+}[W_+(\tau)]
	:=
	\E[W_+(\tau)|\{\mathcal{P}_s\}_{\tau_i \leq s \leq t}]
	=
	W_+(t)\qquad \forall \; t<\tau\in T.
\end{equation}
\par

We will make the additional assumption that the time-reversed process is also a semi-martingale. Hence, we can construct a time reversed filtration $\{\mathcal{F}_\tau\}_{\tau\in T}$, which is a left-continuous and decreasing set of sigma algebras, i.e. $\mathcal{F}_\tau=\cap_{\epsilon>0} \mathcal{F}_{\tau-\epsilon}$ and $\mathcal{F}_{\tau_f}\subseteq\mathcal{F}_s\subseteq\mathcal{F}_t\subseteq\Sigma$  $\forall\,s>t\in T$. Moreover, $X$ is adapted to this filtration and can be decomposed as
\begin{equation}
	X(\tau)= x_f + C_-(\tau) + W_-(\tau),
\end{equation}
where $X(\tau_f)= x_f$, $C_-(\tau_f)=0$ and $W_-(\tau_f)=0$. Furthermore, $W_-$ satisfies the backward martingale property 
\begin{equation}
	\E_{t^-}[W_-(\tau)]:=\E[W_-(\tau)|\{\mathcal{F}_s\}_{t \leq s \leq \tau_f}]=W_-(t)\qquad \forall \; t>\tau\in T.
\end{equation}
\par

For obvious reasons, we will call $\{\mathcal{P}_\tau\}_{\tau\in T}$ the past filtration and $\{\mathcal{F}_\tau\}_{\tau\in T}$ the future filtration. The intersection of the two $\mathfrak{P}_\tau = \mathcal{P}_\tau \cap \mathcal{F}_\tau$, will be called the present sigma algebra, and we denote conditional expectations with respect to this sigma algebra by 
\begin{equation}
	\E_t[X(\tau)]:=\E[X(\tau)|\mathfrak{P}_t].
\end{equation}
Furthermore, we will assume Markovianness of both the forward and backward process, i.e. 
\begin{equation}
	\E_{t^+}[X(\tau)]=\E_{t}[X(\tau)] 
	\qquad {\rm and} \qquad 
	\E_{t^-}[X(\tau)]=\E_{t}[X(\tau)].
\end{equation}
\par

Finally, one can define a sample path for every $\omega\in\Omega$ as the set $\gamma(\omega):=\{X(\tau,\omega):\tau \in T\}$. The measurable space of sample paths is the cylinder $\left(\M^T,{\rm Cyl}(\M^T)\right)$, where we take the cylinder sigma algebra on $\M^T$. This construction allows to interpret the stochastic process as a single random variable $\gamma:(\Omega,\Sigma,\mathbb{P})\rightarrow \left(\M^T,{\rm Cyl}(\M^T)\right)$.

\subsection{Time derivatives}
Stochastic motions are not differentiable, and therefore the notion of velocity is not well defined. However, one can define the conditional velocities for the forward and backward process:
\begin{align}
	v_f^\mu\left[X(\tau),\tau\right] 
	&:= \lim_{h\downarrow 0} \frac{1}{h} \E_{\tau^+} \left[X^\mu(\tau+h) - X^\mu(\tau) \right],\nonumber\\
	v_b^\mu\left[X(\tau),\tau\right] 
	&:= \lim_{h\downarrow 0} \frac{1}{h} \E_{\tau^-} \left[X^\mu(\tau-h) - X^\mu(\tau) \right],
\end{align}
Using these velocities, we can construct the compensators $C_\pm(\tau)$. These c\`adl\`ag processes are given by
\begin{align}
	C_+^\mu(\tau) &= \int_{\tau_i}^{\tau} v_f^\mu (X(s),s) \, ds,\nonumber\\
	C_-^\mu(\tau) &= \int_{\tau}^{\tau_f} v_b^\mu (X(s),s) \, ds.	
\end{align}
Since we are dealing with stochastic processes with non-zero quadratic variation, we can also define 
\begin{align}
	v_f^{\mu\nu} \left[X(\tau),\tau\right] 
	&:= \lim_{h\downarrow 0} \frac{1}{2h} \E_{\tau^+} \Big\{
	[X^\mu(\tau+h) - X^\mu(\tau)] [X^\nu(\tau+h) - X^\nu(\tau)] \Big\},\nonumber\\
	v_b^{\mu\nu} \left[X(\tau),\tau\right] 
	&:= \lim_{h\downarrow 0} \frac{1}{2h} \E_{\tau^-} \Big\{
	[ X^\mu(\tau-h) - X^\mu(\tau) ] [ X^\nu(\tau-h) - X^\nu(\tau) ] \Big\}.
\end{align}
This can be used to construct the compensator\footnote{The compensator of the quadratic variation process is often denoted by the angle bracket $\langle X^\mu, X^\nu\rangle$. We will use $C^{\mu\nu}(\tau)$ instead to avoid confusion with the duality pairing.} $C^{\mu\nu}(\tau)$ of the quadratic variation process $[[X^\mu,X^\nu]]$, which is given by
\begin{align}
	C_+^{\mu\nu}(\tau) &= 2\int_{\tau_i}^{\tau} v_f^{\mu\nu} (X(s),s) ds,\nonumber\\
	C_-^{\mu\nu}(\tau) &= 2\int_{\tau}^{\tau_f} v_b^{\mu\nu} (X(s),s) ds.	
\end{align}
\par 

In practice, we choose the direction of time. We will therefore introduce a slightly modified notion of velocity and define a \textit{forward velocity} and \textit{backward velocity} by
\begin{align}
	v_+(X,\tau) &= v_f(X,\tau),\nonumber\\
	v_-(X,\tau) &= - v_b(X,\tau).
\end{align}
Using the Markov property, these velocities can be defined by\footnote{Note that the backward velocity can equivalently be defined as $v_-^\mu[X(\tau),\tau] := \lim_{h\downarrow 0} \frac{1}{h} \E_{\tau} [X^\mu(\tau) - X^\mu(\tau-h)]$.}
\begin{align}
	v_+^\mu\left[X(\tau),\tau\right] 
	&:= \lim_{h\downarrow 0} \frac{1}{h} \E_{\tau} \left[X^\mu(\tau+h) - X^\mu(\tau) \right],\nonumber\\
	v_-^\mu\left[X(\tau),\tau\right] 
	&:= \lim_{h\uparrow 0} \frac{1}{h} \E_{\tau} \left[X^\mu(\tau+h) - X^\mu(\tau) \right],
\end{align}
and
\begin{align}
	v_+^{\mu\nu} \left[X(\tau),\tau\right] 
	&:= \lim_{h\downarrow 0} \frac{1}{2h} \E_{\tau^+} \Big\{
	[X^\mu(\tau+h) - X^\mu(\tau)] [X^\nu(\tau+h) - X^\nu(\tau)] \Big\},\nonumber\\
	v_-^{\mu\nu} \left[X(\tau),\tau\right] 
	&:= \lim_{h\uparrow 0} \frac{1}{2h} \E_{\tau^-} \Big\{
	[ X^\mu(\tau+h) - X^\mu(\tau) ] [ X^\nu(\tau+h) - X^\nu(\tau) ] \Big\}.
\end{align}
\par 

Reversibility of the process imposes
\begin{equation}
	v_b^{\mu\nu}(\tau) = v_f^{\mu\nu}(\tau),
\end{equation}
and therefore
\begin{equation}
	v_+^{\mu\nu}(\tau) = - v_-^{\mu\nu}(\tau).
\end{equation}
Moreover, the background hypothesis imposes
\begin{align}
	[[X_\mu, X^\nu]](\tau) &= \frac{\hbar}{m} \, \delta_\mu^\nu \, \tau.
\end{align}
Hence,
\begin{align}
	d[[X_\mu, X^\nu]] &= \frac{\hbar}{m} \, \delta_\mu^\nu \, d\tau.
\end{align}
Consequently,
\begin{align}
	v_+^{\mu\nu}[X(\tau),\tau] 
	&= \frac{1}{2 \, d\tau} \E_{\tau}\Big[g^{\mu\rho}(X(\tau))\, d[[X_\rho(\tau), X^\nu(\tau)]]\Big]\nonumber\\
	&= \frac{\hbar}{2m} g^{\mu\nu}(X(\tau)).
\end{align}
\par 

$v_\pm[X(\tau),\tau]$ has the structure of a second order vector, i.e. $v_\pm(x)\in \tilde{T}_x\M$. If the metric is fixed\footnote{In this paper, we only consider test particles in a fixed geometry.}, the second order parts $v_\pm^{\mu\nu}(x)$ are also fixed. The vectors then live in $n$-dimensional subspaces $v_\pm^\mu\in T^\pm_x\M\subset \tilde{T}_x\M$. Since these slices are not invariant under coordinate transformations, we will consider $(\hat{v}_+,\hat{v}_-)\in\hat{T}_x^+\M \oplus \hat{T}_x^-\M$ instead.
\par 

Finally, we define a \textit{current velocity} by
\begin{equation}
	v := \frac{1}{2} \left(v_+ + v_-\right)
\end{equation}	
and an \textit{osmotic velocity} by
\begin{equation}
	u := \frac{1}{2} \left(v_+ - v_-\right).
\end{equation}	
Notice that $v\in T_x\M$ is a first order vector, while $u\in \tilde{T}_x\M$ has the structure of a second order vector.

\subsection{Diffeomorphism invariance}
In classical physics, one imposes a theory to be invariant under diffeomorphisms: general relativity should be invariant under the action of any diffeomorphism $\phi\in C^\infty(\M,\mathcal{N})$. The diffeomorphism $\phi$ induces associated maps on the tangent and cotangent spaces, which are the pullback $\phi^\ast:T^\ast_y\mathcal{N}\rightarrow T^\ast_x\M$ and the pushforward $\phi_\ast:T_x\M\rightarrow T_y\mathcal{N}$, where $y=\phi(x)$. The tangent space and cotangent space are invariant under respectively the pullback and the pushforward.
\par

In quantum physics, we would like to impose the same invariance under diffeomorphisms. However, it is not immediately clear that the $n$-dimensional tangent subspace $\hat{T}_x\M\subset\tilde{T}_x\M$ and cotangent subspace $\hat{T}_x^\ast\M\subset\tilde{T}_x^\ast\M$ with fixed second order parts are invariant spaces under the the pullback $\tilde{\phi}^\ast:\tilde{T}^\ast_y\mathcal{N}\rightarrow \tilde{T}^\ast_x\M$ and pushforward $\tilde{\phi}_\ast:\tilde{T}_x\M\rightarrow \tilde{T}_y\mathcal{N}$ of a diffeomorphism $\phi$. In order to establish this invariance, we require the notion of a Schwartz morphism:\footnote{cf. Definition 6.22 in Ref.~\cite{Emery}.}

\begin{mydef}
	Given two manifolds $\M,\mathcal{N}$ and points $x\in\M$, $y\in\mathcal{N}$, a linear mapping $f:\tilde{T}_x\M\rightarrow\tilde{T}_y\mathcal{N}$ is called a \textnormal{Schwartz morphism}, if
	\begin{enumerate}
		\item $f(T_x\M)\subset T_y\mathcal{N}$,
		\item $\forall\, L\in \tilde{T}_x\M, \; \mathcal{H}^{\ast}(f(L))= (f^\circ\otimes f^\circ)\mathcal{H}^{\ast}(L)$,
	\end{enumerate}
	where $f^\circ$ is the restriction of $f$ to $T_x\M$.
\end{mydef}

A Schwartz morphism is thus a morphism that leaves the slices $\hat{T}_x\M$ invariant. Furthermore, it can be shown\footnote{cf. Exercise 6.23 in Ref.~\cite{Emery}} that a mapping $f:\tilde{T}_x\M\rightarrow\tilde{T}_y\mathcal{N}$ is a Schwartz morphism if and only if $f=\tilde{T}_x\phi$ for a smooth $\phi:\M\rightarrow\mathcal{N}$ with $\phi(x)=y$. It immediately follows that the pushforward $\tilde{\phi}_\ast$ of a diffeomorphism $\phi$ is a Schwartz morphism. Therefore, all slices $\hat{T}\M\subset\tilde{T}\M$ are invariant under the pushforward $\tilde{\phi}_\ast:\tilde{T}_x\M\rightarrow\tilde{T}_{\phi(x)}\mathcal{N}$ induced by a diffeomorphism $\phi:\M\rightarrow\mathcal{N}$. Moreover, all slices $\hat{T}^\ast\M\subset\tilde{T}^\ast\M$ are invariant under the pullback $\tilde{\phi}^\ast:\tilde{T}^\ast_{\phi(x)}\mathcal{N}\rightarrow\tilde{T}^\ast_x\M$ of the diffeomorphism $\phi$. We note that this invariance is a consequence of the construction of the `covariant slices' $\hat{T}_x\M$.

\section{Integration Along Semi-Martingales}\label{sec:integration}
In the previous sections, we have introduced manifold valued semi-martingales and second order geometry. This allows us to construct a notion of integration along semi-martingales on manifolds. This section is loosely based on the review by Emery \cite{Emery}. For mathematical detail we refer to this work by Emery \cite{Emery} or the original works by Schwartz \cite{Schwartz} and Meyer \cite{Meyer}.
\par

In first order geometry, one defines integrals using forms $\omega\in T^\ast\M$. The integral of a form along a curve $\gamma:I\rightarrow\M$ with $I\subset\R$ is given by
\begin{equation}
	\int_\gamma:T^\ast\M\rightarrow\R
	\qquad {\rm s.t.} \qquad
	\omega \mapsto \int_\gamma \omega(x),
\end{equation}
which can be written as
\begin{equation}
	\int_\gamma \omega
	= \int_{\tau_i}^{\tau_f} \omega_\mu \, d \gamma^\mu
	= \int_{\tau_i}^{\tau_f} \omega_\mu \dot{\gamma}^\mu \, d \tau,
\end{equation}
where $d\gamma=\gamma^\ast(\omega)$. 
If we assume that the form can be written as a differential form $\omega=d F$ for a function $F\in C^\infty(\M,\R)$ we find
\begin{equation}
	\int_\gamma d F(x)
	= \int_{\tau_i}^{\tau_f} \p_\mu F(\gamma)\, d \gamma^\mu
	= \int_{\tau_i}^{\tau_f} \p_\mu F(\gamma) \dot{\gamma}^\mu \, d \tau.
\end{equation}
Moreover, the fundamental theorem for line integrals states
\begin{equation}
	\int_\gamma d F(x) = F[\gamma(\tau_f)] - F[\gamma(\tau_i)].
\end{equation}
\par

One can analogously construct an integral of second order forms $\Omega\in \tilde{T}^\ast\M$. The integral of a second order form along a semi-martingale $X$ can be written as
\begin{equation}
	\int_X:\tilde{T}^\ast\M\rightarrow\R
	\qquad {\rm s.t.} \qquad
	\Omega \mapsto \int_X \Omega(x)
\end{equation}
with
\begin{align}
	\int_X \Omega
	&= \int_{\tau_i}^{\tau_f} \omega_\mu \, d_2X^\mu 
	+ \int_{\tau_i}^{\tau_f} \omega_{\mu\nu} \, dX^\mu \cdot dX^\nu\nonumber\\
	&= \int_{\tau_i}^{\tau_f} \hat{\omega}_\mu \, d_2 \hat{X}^\mu 
	+ \int_{\tau_i}^{\tau_f} \hat{\omega}_{\mu\nu} \, d \hat{X}^\mu \cdot d \hat{X}^\nu,
\end{align}
where $(d_2X\quad dX dX)=X^\ast(\Omega)$.
If we assume that the form can be written as a differential form $\Omega=d_2 F$ for a function $F\in C^\infty(\M,\R)$, we find
\begin{align}
	\int_X d_2 F(x)
	&= \int_{\tau_i}^{\tau_f} \p_\mu F(X)\, d_2 X^\mu
	+ \int_{\tau_i}^{\tau_f} \p_\mu\p_\nu F(X)\, d X^\mu \cdot d X^\nu \nonumber\\
	&= \int_{\tau_i}^{\tau_f} \nabla_\mu F(X)\, d_2 \hat{X}^\mu
	+ \int_{\tau_i}^{\tau_f} \nabla_\mu\nabla_\nu F(X)\, d \hat{X}^\mu \cdot d \hat{X}^\nu.
\end{align}
The fundamental theorem for line integrals can be extended to the second order context, such that\footnote{cf. Theorem 6.24 in Ref.~\cite{Emery}.}
\begin{equation}
	\int_X d_2 F(x) = F[X(\tau_f)] - F[X(\tau_i)],
\end{equation}
\par

Moreover, one can relate the second order integral to first order order integrals. For this we consider a form $\omega\in T\M\subset\tilde{T}\M$. We can then construct two second order integrals, that are manifestly invariant under coordinate transformations, using the maps $\underline{d}$ and $\mathcal{G}$ respectively:
\begin{align}
	\dashint_X \omega 
	&= \int_X \underline{d}\omega \nonumber\\
	&= \int_{\tau_i}^{\tau_f} \omega_\mu\, d_2X^\mu
	+ \int_{\tau_i}^{\tau_f} \p_\nu \omega_\mu\, dX^\mu \cdot dX^\nu\nonumber\\
	&= \int_{\tau_i}^{\tau_f} \hat{\omega}_\mu\, d_2 \hat{X}^\mu
	+ \int_{\tau_i}^{\tau_f} \nabla_\nu \omega_\mu\, d \hat{X}^\mu \cdot d \hat{X}^\nu, \label{eq:StratonovichFirstForm}
\end{align}
and
\begin{align}
	\lowint_X \omega 
	&= 
	\int_X \mathcal{G}(\omega)\nonumber\\
	&= 
	\int_{\tau_i}^{\tau_f} \omega_\mu d_2 X^\mu
	+ \int_{\tau_i}^{\tau_f} \omega_\mu \Gamma^\mu_{\nu\rho}\, dX^\nu \cdot dX^\rho \nonumber\\
	&= 
	\int_{\tau_i}^{\tau_f} \hat{\omega}_\mu\, d_2 \hat{X}^\mu.\label{eq:ItoFirstForm}
\end{align}
The first of these integrals is a Stratonovich integral,\footnote{cf. Definition 7.3 and Proposition 7.4 in Ref.~\cite{Emery}.} while the second is an It\^o integral.\footnote{cf. Definition 7.33 and Proposition 7.34 in Ref.~\cite{Emery}.} We immediately find a relation between the two
\begin{equation}
	\dashint_X \omega 
	= \lowint_X \omega 
	+ \int_{\tau_i}^{\tau_f} \nabla_\nu \hat{\omega}_\mu\,  d \hat{X}^\mu \cdot d \hat{X}^\nu.
\end{equation}
In order to evaluate the integral over the second order part we use that the integral over a bilinear form is given by\footnote{cf. Theorem 3.8 in Ref.~\cite{Emery}.}
\begin{equation}\label{eq:IntBiForm}
	\int_{\tau_i}^{\tau_f} f_{\mu\nu}(X,\tau)\, dX^\mu \otimes dX^\nu
	=
	\int_{\tau_i}^{\tau_f} f_{\mu\nu}(X,\tau) \, d[[X^\mu, X^\nu]].
\end{equation}
Using the map $\mathcal{H}$ one can then map the integral over the second order part to an integral over a bilinear form. This yields\footnote{cf. Proposition 6.31 in Ref.~\cite{Emery}.}
\begin{equation}
	\int_{\tau_i}^{\tau_f} f_{\mu\nu}(X,\tau)\, dX^\mu \cdot dX^\nu
	=
	\frac{1}{2} \int_{\tau_i}^{\tau_f} f_{\mu\nu}(X,\tau) \, d[[X^\mu, X^\nu]]
	=
	\int_{\tau_i}^{\tau_f} f_{\mu\nu}(X,\tau)\, v^{\mu\nu}(X,\tau)\, d\tau.
\end{equation}
\par 

Moreover, if $\omega$ can be written as a differential form $\omega=dF$, the two first order integrals can be written as\footnote{We use the notation $\lowint f_\mu(X)\, d_+ \hat{X}^\mu$ instead of $\lowint f_\mu(X)\, d X^\mu$ to make the covariance of the expression explicit.}
\begin{align}
	\int_X d_2 F(x) 
	&= \dashint_{\tau_i}^{\tau_f} \p_\mu F(X)\, dX^\mu,\nonumber\\
	\int_X d_2 F(x) 
	&= \lowint_{\tau_i}^{\tau_f} \nabla_\mu F(X) \, d_+ \hat{X}^\mu + \int_{\tau_i}^{\tau_f} \nabla_{\mu} \nabla_{\nu} F(X) \, d \hat{X}^\mu\cdot d \hat{X}^\nu.
\end{align}
Using the decomposition of the semi-martingale, we can then write
\begin{align}
	\int_X d_2 F(x) 
	&= \int_{\tau_i}^{\tau_f} v^\mu(X,\tau) \p_\mu F(X)\, d\tau
	+ \dashint_{\tau_i}^{\tau_f} \p_\mu F(X)\, dW^\mu,\\
	\int_X d_2 F(x) 
	&= \int_{\tau_i}^{\tau_f} \hat{v}_+^\mu(X,\tau) \nabla_\mu F(X)\, d\tau 
	+ \lowint_{\tau_i}^{\tau_f} \nabla_\mu F(X) \, dW_+^\mu
	+ \int_{\tau_i}^{\tau_f} \hat{v}_+^{\mu\nu}(X,\tau) \nabla_{\mu} \nabla_{\nu} F(X) \, d\tau.\nonumber
\end{align}
Notice that all integrals are manifestly invariant under coordinate transformations. Furthermore, the It\^o integral is a local martingale, i.e.
\begin{equation}\label{eq:MartingaleProperty}
	\E_{\tau_i^+} \left[ \lowint_{\tau_i}^{\tau} \nabla_\mu F(X) \, dW_+^\mu \right] = 0.
\end{equation}
\par

In addition, we will construct a backward It\^o integral such that
\begin{align}
	\int_X d_2 F(x) 
		&= \upint_{\tau_i}^{\tau_f} \nabla_\mu F(X) \, d_- \hat{X}^\mu - \int_{\tau_i}^{\tau_f} \nabla_{\mu} \nabla_{\nu} F(X) \, d \hat{X}^\mu \cdot d \hat{X}^\nu \\
		&= \int_{\tau_i}^{\tau_f}  \hat{v}_-^\mu(X,\tau) \nabla_\mu F(X)\, d\tau 
		+ \upint_{\tau_i}^{\tau_f} \nabla_\mu F(X) \, dW_-^\mu
		+ \int_{\tau_i}^{\tau_f} \hat{v}_-^{\mu\nu}(X,\tau) \nabla_{\mu} \nabla_{\nu} F(X)\,  d\tau.\nonumber
\end{align}
The backward integral is a local backward martingale, i.e.
\begin{equation}
\E_{\tau_f^-} \left[ \upint_{\tau}^{\tau_f} \nabla_\mu F(X) \, dW_-^\mu \right] = 0.
\end{equation}
We note that the three integrals are related by
\begin{equation}
	\dashint_X dF(x) = \frac{1}{2} \left( \lowint_X dF(x) + \upint_X dF(x) \right).
\end{equation}
\par 

Let us now relate the Stratonovich and It\^o integral to their well known definitions in $\R^n$. If there exists a coordinate chart $\chi:U\rightarrow \R^n$ such that $f([\tau_i,\tau_f])\subset U$, we have\footnote{This is a consequence of Theorem 7.14 and Theorem 7.37 in Ref.~\cite{Emery}.}
\begin{align}
	\dashint_{\tau_i}^{\tau_f} f_{\mu}(X,\tau)\, d X^\mu 
	&:=
	\lim_{k\rightarrow\infty} \sum_{[\tau_{j},\tau_{j+1}]\in \pi_k} \frac{1}{2} \Big[
	f_\mu \big(X(\tau_j),\tau_j\big) + f_\mu \big(X(\tau_{j+1}),\tau_{j+1}\big)\Big]\nonumber\\
	& \qquad  \qquad \qquad \qquad 
	\times \Big[ X^\mu(\tau_{j+1}) - X^\mu(\tau_{j})\Big],\nonumber\\
	\lowint_{\tau_i}^{\tau_f} f_{\mu}(X,\tau)\, d_+ X^\mu 
	&:=
	\lim_{k\rightarrow\infty} \sum_{[\tau_{j},\tau_{j+1}]\in \pi_k} f_\mu \big(X(\tau_j),\tau_j\big) \Big[ X^\mu(\tau_{j+1}) - X^\mu(\tau_{j})\Big],\nonumber\\
	\upint_{\tau_i}^{\tau_f} f_{\mu}(X,\tau)\, d_- X^\mu 
	&:=
	\lim_{k\rightarrow\infty} \sum_{[\tau_{j},\tau_{j+1}]\in \pi_k} f_\mu \big(X(\tau_{j+1}),\tau_{j+1}\big) \Big[ X^\mu(\tau_{j+1}) - X^\mu(\tau_{j})\Big],\nonumber\\
	\int_{\tau_i}^{\tau_f} f_{\mu\nu}(X,\tau)\, d[[X^\mu,X^\nu]] 
	&:=
	\lim_{k\rightarrow\infty} \sum_{[\tau_{j},\tau_{j+1}]\in \pi_k} f_{\mu\nu} \big(X(\tau_{j}),\tau_{j}\big) \Big[ X^\mu(\tau_{j+1}) - X^\mu(\tau_{j})\Big]\nonumber\\
	& \qquad  \qquad \qquad \qquad
	\times \Big[ X^\nu(\tau_{j+1}) - X^\nu(\tau_{j})\Big],\label{eq:IntSumRep}
\end{align}
where $\pi_k$ is a partition of $[\tau_i,\tau_f]$, $f_\mu=(\chi \circ f)_\mu$ and $X^\mu=(\chi \circ X)^\mu$. We thus have
\begin{align}
	\dashint_{\tau_i}^{\tau_f} f_{\mu}(X,\tau)\, d X^\mu 
	&=
	\frac{1}{2} \left(  \lowint_{\tau_i}^{\tau_f} f_{\mu}(X,\tau)\, d_+ X^\mu 
	+ \upint_{\tau_i}^{\tau_f} f_{\mu}(X,\tau)\, d_- X^\mu\right),
\end{align}
and we will define an osmotic integral by
\begin{align}
	\dashint_{\tau_i}^{\tau_f} f_{\mu}(X,\tau)\, d_\circ X^\mu 
	:=
	\frac{1}{2} \left(  \lowint_{\tau_i}^{\tau_f} f_{\mu}(X,\tau)\, d_+ X^\mu 
	- \upint_{\tau_i}^{\tau_f} f_{\mu}(X,\tau)\, d_- X^\mu\right).
\end{align}

\subsection{Integration by parts}
In this subsection, we state two integration by parts formulae, that will be useful for stochastic variational calculus. The first is given by
\begin{align}
	\int_{\tau_i}^{\tau_f} d\left[ f_\mu(\tau) g^\mu(\tau) \right]
	&=
	f_\mu(\tau_f)\, g^\mu(\tau_f) 
	- f_\mu(\tau_i)\, g^\mu(\tau_i) \nonumber\\
	&=
	\dashint_{\tau_i}^{\tau_f} f_\mu(\tau)\, d g^\mu(\tau) 
	+ \dashint_{\tau_i}^{\tau_f} g^\mu(\tau)\, d f_\mu(\tau)\nonumber\\
	&=
	\lowint_{\tau_i}^{\tau_f} f_\mu(\tau) \, d_+ 
	g^\mu(\tau) + \lowint_{\tau_i}^{\tau_f} g^\mu(\tau) \, d_+ f_\mu(\tau) 
	+ 2 \int_{\tau_i}^{\tau_f} d f_\mu(\tau) \cdot d g^\mu(\tau) \nonumber\\
	&=
	\upint_{\tau_i}^{\tau_f} f_\mu(\tau) \, d_- g^\mu(\tau) 
	+ \upint_{\tau_i}^{\tau_f} g^\mu(\tau) \, d_- f_\mu(\tau) 
	- 2 \int_{\tau_i}^{\tau_f} d f_\mu(\tau) \cdot d g^\mu(\tau),
\end{align}
where we write $f_\mu(\tau)= f_\mu(X(\tau),\tau)$, $g^\mu(\tau)= g^\mu(X(\tau),\tau)$.
We immediately find
\begin{equation}\label{eq:OsmoticIntegral}
	\int f_\mu(\tau) \, d_\circ g^\mu(\tau) + \int g^\mu(\tau) \, d_\circ f_\mu(\tau) 
	=
	- 2 \int df_\mu(\tau) \cdot dg^\mu(\tau),
\end{equation}
where we recall
\begin{equation}
	\int df_\mu(\tau) \cdot dg^\mu(\tau)  = \frac{1}{2} \int d[[f_\mu,g^\mu]](\tau).
\end{equation}
\par 

There exists another integration by parts formula, which can be derived from eq.~\eqref{eq:IntSumRep} and is given by\footnote{See also e.g. Refs.\cite{Nelson,Zambrini} for a derivation of this formula}
\begin{align}\label{eq:PIOther}
	\int_{\tau_i}^{\tau_f} d\left[ f_\mu(\tau) g^\mu(\tau) \right]
	&=
	\lowint f_\mu(\tau) \, d_+ g^\mu(\tau) + \upint g^\mu(\tau) \, d_- f_\mu(\tau)\nonumber\\
	&=
	\upint f_\mu(\tau) \, d_- g^\mu(\tau) + \lowint g^\mu(\tau) \, d_+ f_\mu(\tau).
\end{align}
Combining eqs. \eqref{eq:OsmoticIntegral} and \eqref{eq:PIOther} then yields
\begin{equation}\label{eq:OsmoticIntegral2}
	\int f_\mu(\tau) \, d_\circ g^\mu(\tau) 
	=
	\int g^\mu(\tau) \, d_\circ f_\mu(\tau) 
	=
	- \int df_\mu(\tau) \cdot dg^\mu(\tau).
\end{equation}
\section{Stochastic Variational Calculus}\label{sec:StochVarCalc}
In this section, we discuss stochastic variational calculus as developed by Yasue \cite{Yasue,Yasue:1981wu,Zambrini}. We will consider the tangent bundle 
\begin{equation}
	\hat{T}\M = \bigsqcup_{x\in\M} \left(\hat{T}_x^+\M \oplus \hat{T}_x^-\M \right),
\end{equation}
which can be endowed with a $(3n)$-dimensional manifold structure with coordinates $(x^\mu,v_+^\mu,v_-^\mu)$. We define the Lagrangian as a map
\begin{equation}\label{eq:LagrangianParticle}
	L: \hat{T}\M \rightarrow \R,
\end{equation} 
and the action as the integral
\begin{equation}
	S = \E \left[\int_{\tau_i}^{\tau_f} L(X,V_+,V_-) \, d\tau \right].
\end{equation}
Equivalently the action can be expressed as a function of the processes $X$, $V(V_+,V_-)$ and $U(V_+,V_-)$, which we will use later on. We emphasize that $V_\pm(\tau)$ are processes on the tangent bundle, while $v_\pm(X,\tau)$ are second order vector fields. The two are related as follows
\begin{align}
	\lim_{s \rightarrow \tau} \E_\tau\left[ V_+^\mu(s) \right] &= v_+^\mu(X,\tau),\nonumber\\
	\lim_{s \rightarrow \tau} \E_\tau\left[ V_-^\mu(s) \right] &= v_-^\mu(X,\tau).
\end{align}
\par

As we intend to do variational calculus, we require the notion of a norm on the space of manifold valued time-reversible semi-martingales.  In order to construct such a norm, we would like to split the space of all processes into spaces of time-like, space-like, and null-like processes. For this, we need to define the notion of a time-like process. 
We will call the process $X=X(\tau)$ \textit{time-like}, if
\begin{equation}
	g_{\mu\nu}(X) \, v^\mu(X,\tau) \, v^\nu(X,\tau) < 0 \qquad \forall\;\tau\in T.
\end{equation}
Moreover, we call the process \textit{space-like}, if
\begin{equation}
	g_{\mu\nu}(X) \, v^\mu(X,\tau) \, v^\nu(X,\tau) > 0 \qquad \forall\;\tau\in T,
\end{equation}
and \textit{light-like} or \textit{null-like}, if 
\begin{equation}
	g_{\mu\nu}(X) \, v^\mu(X,\tau) \, v^\nu(X,\tau) = 0 \qquad \forall\;\tau\in T.
\end{equation}
\par

Note that sample paths of a time-like process are not necessarily time-like. Indeed, for a time-like process we have
\begin{equation}
	\E\Big[g_{\mu\nu}(X(\tau))\, dX^\mu(\tau) \otimes dX^\nu(\tau) \Big] < 0 \qquad \forall\;\tau\in T.
\end{equation}
However, this relation does not hold without the expectation value. Therefore, sample paths can contain segments that are not time-like. A similar remark holds for space-like and light-like processes.
\par
 
We will now restrict the semi-martingales on $\M$ to those that are time-like. After a Wick rotation, the space of these time-like processes can be equipped with the $L^2$-norm
\begin{equation}\label{eq:L2orm}
	||X||= \sqrt{ \E\left[\int \big|X_\mu(\tau) X^\mu(\tau) \big| \, d\tau \right] },
\end{equation}
which is the conventional choice in quantum mechanics.

\subsection{Euler-Lagrange equations}
The stochastic Euler Lagrange equations can be derived similar to the classical Euler-Lagrange equations. We vary the action with respect to a semi-martingale $\delta X$ independent of $X$ that satisfies
\begin{equation}
	\delta X(\tau_i)=\delta X(\tau_f)=0.
\end{equation}
This leads to
\begin{align}
	\delta S(X) 
	&:= 
	S(X+\delta X) - S(X)\nonumber\\
	&= 
	\E \left[\int_{\tau_i}^{\tau_f} L\left(X+\delta X, V_+ + \delta V_+, V_- + \delta V_- \right) \, d\tau \right]
	- \E \left[\int_{\tau_i}^{\tau_f} L\left(X,V_+,V_-\right) \, d\tau \right] \nonumber\\
	&=
	\E \left[\int_{\tau_i}^{\tau_f} \left\{ 
	\frac{\p L(X, V_+, V_- )}{\p X^\mu} \delta X^\mu
	+ \frac{\p L(X, V_+, V_- )}{\p V_+^\mu} \delta V_+^\mu
	\right. \right. \nonumber\\
	&\qquad \qquad \qquad \qquad \qquad \qquad \left. \left.
	+ \frac{\p L(X, V_+, V_- )}{\p V_-^\mu} \delta V_-^\mu \right\} d\tau \right] 
	+ \mathcal{O}(||\delta X||^2)\nonumber\\
	&=
	\E \left[\int_{\tau_i}^{\tau_f} \left\{ 
	\frac{\p L(X, V_+, V_- )}{\p X^\mu} \delta X^\mu d\tau
	+ \frac{\p L(X, V_+, V_- )}{\p V_+^\mu} d_+ \delta X^\mu
	\right. \right. \nonumber\\
	&\qquad \qquad \qquad \qquad \qquad \qquad \left. \left.
	+ \frac{\p L(X, V_+, V_- )}{\p V_-^\mu} d_- \delta X^\mu \right\} \right] 
	+ \mathcal{O}(||\delta X||^2)\nonumber\\
	&=
	\E \left[\int_{\tau_i}^{\tau_f} \delta X^\mu \left\{ 
	\frac{\p L(X, V_+, V_- )}{\p X^\mu} d\tau
	- d_- \frac{\p L(X, V_+, V_- )}{\p V_+^\mu} 
	\right. \right. \nonumber\\
	&\qquad \qquad \qquad \qquad \qquad \qquad \qquad \qquad \left. \left.
	- d_+  \frac{\p L(X, V_+, V_- )}{\p V_-^\mu}
	\right\} \right] 
	+ \mathcal{O}(||\delta X||^2),
\end{align}
where we used the partial integration formula \eqref{eq:PIOther}. We find a system of stochastic differential equations given by
\begin{equation}\label{eq:StochEL}
	\int_{\tau_i}^{\tau_f} \frac{\p}{\p X^\mu} L(X, V_+, V_- ) d\tau
	=
	\int_{\tau_i}^{\tau_f} \left\{ 
	d_- \frac{\p }{\p V_+^\mu} L(X, V_+, V_- ) 
	+ d_+  \frac{\p}{\p V_-^\mu} L(X, V_+, V_- ) \right\}
\end{equation}
or equivalently
\begin{equation}\label{eq:StochEL2}
	\int_{\tau_i}^{\tau_f} \frac{\p}{\p X^\mu} L(X, V, U ) d\tau
	=
	\int_{\tau_i}^{\tau_f} \left\{ 
	d \frac{\p }{\p V^\mu} L(X, V, U ) 
	- d_\circ  \frac{\p}{\p U^\mu} L(X, V, U ) \right\}.
\end{equation}
Since $\delta X \indep X$, the osmotic integral vanishes, and we obtain
\begin{equation}\label{eq:StochEL3}
	\int_{\tau_i}^{\tau_f} \frac{\p}{\p X^\mu} L(X, V, U ) d\tau
	=
	\int_{\tau_i}^{\tau_f} 
	d \frac{\p }{\p V^\mu} L(X, V, U ).
\end{equation}

\subsection{Hamilton equations}
As in classical physics, one can define an Hamiltonian picture.
We define the generalized momenta by
\begin{align}
	P^+_\mu(\tau) &= \frac{\p L}{\p V_+^\mu},\nonumber\\
	P^-_\mu(\tau) &= \frac{\p L}{\p V_-^\mu}.
\end{align}
and the Hamiltonian as the Legendre transform
\begin{equation}
	H(X,P^+,P^-) = P^+_\mu V_+^\mu + P^-_\mu  V_-^\mu - L(X,V_+,V_-).
\end{equation}
We can take a first order total derivative. This yields
\begin{equation}
	dH 
	= \frac{\p H}{\p X^\mu} dX^\mu 
	+ \frac{\p H}{\p P^+_\mu} dP^+_\mu
	+ \frac{\p H}{\p P^-_\mu} dP^-_\mu
\end{equation}
and
\begin{align}
	dH 
	&= P^+_\mu d V_+^\mu +  V_+^\mu d P^+_\mu
	+ P^-_\mu d V_-^\mu +  V_-^\mu d P^-_\mu 
	- \frac{\p L}{\p X^\mu} dX^\mu - \frac{\p L}{\p V_+^\mu} dV_+^\mu 
	- \frac{\p L}{\p V_-^\mu} dV_-^\mu \nonumber\\
	&= V_+^\mu d P^+_\mu +  V_-^\mu d P^-_\mu 
	- \left( \frac{d_-}{d\tau} P^+_\mu + \frac{d_+}{d\tau} P^-_\mu \right) dX^\mu.
\end{align}
One can then read off the Hamilton equations:
\begin{align}
	V_+^\mu(\tau) &= \frac{\p H}{\p P^+_\mu},\nonumber\\
	V_-^\mu(\tau) &= \frac{\p H}{\p P^-_\mu}
\end{align}
and
\begin{equation}
	\int\left( d_+ P^-_\mu + d_- P^+_\mu \right) = - \int \frac{\p H}{\p X^\mu} d\tau.
\end{equation}
Furthermore, if an explicit proper time dependence is introduced, one finds
\begin{equation}
	\frac{\p}{\p \tau} H(X,P^+,P^-,\tau) = - \frac{\p}{\p \tau} L(X,V_+,V_-,\tau).
\end{equation}
\par

As is the case for the Lagrangian, one can express the Hamiltonian in terms of \textit{current} and \textit{osmotic momenta}. These can be defined as
\begin{align}
	P_\mu(\tau) &= \frac{\p}{\p V^\mu} L(X,V,U),\nonumber\\
	Q_\mu(\tau) &= \frac{\p}{\p U^\mu} L(X,V,U).
\end{align}
The Hamiltonian is then given by
\begin{equation}
	H(X,P,Q) = P_\mu V^\mu + Q_\mu U^\mu - L(X,V,U).
\end{equation}
This leads to the Hamilton equations
\begin{align}
	V^\mu(\tau) &= \frac{\p H}{\p P_\mu},\nonumber\\
	U^\mu(\tau) &= \frac{\p H}{\p Q_\mu}.
\end{align}
and
\begin{equation}
	\int d P_\mu = - \int \frac{\p H}{\p X^\mu} d\tau.
\end{equation}
Let us summarize the relation between $U,V,V_+,V_-$:
\begin{align}
	V = \frac{1}{2} \left( V_+ + V_-\right),
	\qquad \qquad
	& V_+ = V + U, \nonumber
	\\
	U = \frac{1}{2} \left( V_+ - V_-\right),
	\qquad \qquad
	& V_- = V - U.
\end{align}
Furthermore, for $P,Q,P_+,P_-$ we have
\begin{align}
	P = P_+ + P_-,
	\qquad \qquad
	& P_+ = \frac{1}{2} \left(P + Q \right),\nonumber
	\\
	Q = P_+ - P_-,
	\qquad \qquad
	& P_- = \frac{1}{2} \left(P - Q \right).
\end{align}
\subsection{Hamilton-Jacobi equations}\label{sec:HamJac}
The Hamilton-Jacobi equations play an important role in the derivation of the Schr\"odinger equation in stochastic quantization. We will therefore review the derivation of these equations. We define Hamilton's principal function as the action conditioned on its end point
\begin{equation}
	S(X,\tau) = \E\left[ \int_{\tau_i}^{\tau} L(X,V_+,V_-)\, ds \Big| X(\tau) \right],
\end{equation}
such that the Euler-Lagrange equations are satisfied.
\par

We consider the variation of the principal function under a variation of the end point. This yields
\begin{align}
	\delta S(X,\tau)
	&= S(X+\delta X,\tau) - S(X,\tau) \nonumber\\
	&= \E\left[ \int_{\tau_i}^{\tau} L(X,V_+,V_-)\, ds \Big| X(\tau)+\delta X(\tau) \right]
	- \E\left[ \int_{\tau_i}^{\tau} L(X,V_+,V_-)\, ds \Big| X(\tau) \right]\nonumber\\
	&= \E\left[ \int_{\tau_i}^{\tau} L(X+\delta X,V_+ + \delta V_+ ,V_- + \delta V_-) \, ds 
	- \int_{\tau_i}^{\tau} L(X,V_+,V_-) \, ds \Big| X(\tau), \delta X(\tau) \right]\nonumber\\
	&= \E\left[ \int_{\tau_i}^{\tau} \left\{ 
	\frac{\p}{\p X^\mu} L(X,V_+ ,V_- ) \, \delta X^\mu
	+\frac{\p}{\p V_+^\mu} L(X,V_+ ,V_- ) \, \delta V_+^\mu
	\right. \right. \nonumber\\
	& \qquad \qquad \left. \left.
	+\frac{\p}{\p V_-^\mu} L(X,V_+ ,V_- )\, \delta V_-^\mu \right\} ds
	\Big| X(\tau), \delta X(\tau) \right]
	+ \mathcal{O}\left(||\delta X||^2\right) \nonumber\\
	&= \E\left[ \int_{\tau_i}^{\tau} \left\{ 
	\delta X^\mu \, d_- \frac{\p L}{\p V_+^\mu}
	+ \delta X^\mu \, d_+ \frac{\p L}{\p V_-^\mu}
	\right. \right. \nonumber\\
	& \qquad \qquad \left. \left.
	+ \frac{\p L}{\p V_+^\mu} \, d_+ \delta X^\mu
	+ \frac{\p L}{\p V_-^\mu}  \, d_- \delta X^\mu  \right\}
	\Big| X(\tau), \delta X(\tau) \right]
	+ \mathcal{O}\left(||\delta X||^2\right) \nonumber\\
	&= \E\left[ \int_{\tau_i}^{\tau}  
	d \left[ \left( \frac{\p L}{\p V_+^\mu} + \frac{\p L}{\p V_-^\mu}  \right) \delta X^\mu \right]
	+ \mathcal{O}\left(||\delta X||^2\right)  	
	\Big| X(\tau), \delta X(\tau) \right]\nonumber\\
	&= \Big(p^+_\mu(X,\tau) + p^-_\mu(X,\tau)  \Big) \delta X^\mu 
	+ \mathcal{O}\Big(||\delta X||^2\Big),
\end{align}
where we used the Euler-Lagrange equations in the fifth line. Furthermore, in the third line, we have rewritten the original trajectory which is the minimal path between $(\tau_i,x_i)$ and $(\tau,X(\tau)+\delta X(\tau))$ as two independent trajectories $X,\delta X$, which are the minimal paths between $(\tau_i,x_i)$ and $(\tau,X(\tau))$ and between $(\tau_i,0)$ and $(\tau,\delta X(\tau))$ respectively. 
\par 

We conclude with the first Hamilton-Jacobi equation
\begin{equation}\label{eq:HamJac1}
	\nabla_\mu S(X,\tau) = p^+_\mu(X,\tau) + p^-_\mu(X,\tau) = p_\mu(X,\tau).
\end{equation}
Moreover, taking a first order total derivative of Hamilton's principal function yields
\begin{align}
	dS &= \E_\tau\left[L\, d\tau\right],\nonumber\\
	dS &= \E_\tau\left[\frac{\p S}{\p x^\mu} dX^\mu + \frac{\p S}{\p \tau} d\tau\right].
\end{align}
This leads to the second Hamilton-Jacobi equation
\begin{equation}\label{eq:HamJac2}
	\frac{\p }{\p \tau} S(X,\tau)
	= \E_\tau\left[L(X,V,U)\right] - p_\mu v^\mu.
\end{equation}

\subsection{Kolmogorov equations}
In this section, we derive the Kolmogorov equations. Although these do not follow from a variational principle, they are another crucial ingredient for the derivation of the Schr\"odinger equation.
\par 

Let $\mu(x,\tau)$ be a probability measure on $\M\times T$, such that
\begin{equation}
	\int_{\M\times T} f(x,\tau)\, d\mu(x,\tau) = \int_T \E\left[f(X(\tau),\tau) \right] d\tau
\end{equation}
for any smooth function $f$ compactly supported on $\M\times \rm{int}(T)$, where $\rm{int}(T)$ is the interior of $T$. We will assume that the probability density $\rho$ associated to the measure $\mu$ exists, such that $d\mu(x,\tau)=\sqrt{|g|}\rho(x,\tau)d^n x d\tau$. Then
\begin{align}
	0 
	&= \E[f(X(\tau_f),\tau_f)] - \E[[f(X(\tau_i),\tau_i)] \nonumber\\
	&= \int_T \frac{d_2}{d\tau} \E[f(X(\tau),\tau)] d\tau \nonumber\\
	&= \int_T \E\left[\frac{d_2}{d\tau} f(X(\tau),\tau)\right] d\tau \nonumber\\
	&= \int_T \E\left[\E_\tau\left[ \frac{d_2}{d\tau} f(X(\tau),\tau)\right]\right] d\tau \nonumber\\
	&= \int_T \E\left[\left( \frac{\partial}{\partial \tau} + \hat{v}^\mu(X,\tau) \nabla_\mu + \hat{v}^{\mu\nu}(X,\tau) \nabla_\mu \nabla_\nu \right) f\left(X,t\right)\right] d\tau \nonumber\\
	&= \int_{\mathcal{M}\times T} \left( \frac{\partial}{\partial \tau} + \hat{v}^\mu(x,\tau) \nabla_\mu +  \hat{v}^{\mu\nu}(x,\tau) \nabla_\mu \nabla_\nu \right) f(x,\tau) \, d\mu(x,\tau) \nonumber\\
	&= \int_{\mathcal{M}\times T} \sqrt{|g|}\, \rho(x,\tau) \left( \frac{\partial}{\partial \tau} + \hat{v}^\mu(x,\tau) \nabla_\mu +  \hat{v}^{\mu\nu}(x,\tau) \nabla_\mu \nabla_\nu \right) f(x,\tau) \, d^n x \, d\tau \nonumber\\
	&= \int_{\mathcal{M}\times T} \sqrt{|g|}\,  f(x,\tau) \left(- \frac{\partial}{\partial \tau} \rho(x,\tau) - \nabla_\mu \left[ \hat{v}^\mu(x,\tau)\, \rho(x,\tau) \right] + \nabla_\mu \nabla_\nu \left[ \hat{v}^{\mu\nu}(x,\tau)\, \rho(x,\tau) \right]  \right) d^nx \, d\tau
\end{align}
for all compactly supported functions $f$. We can choose $v=v_\pm$, and plug in the background hypothesis
\begin{equation}
	\hat{v}_\pm^{\mu\nu} = \pm \frac{\hbar}{2m} g^{\mu\nu}.
\end{equation}
This leads to the Kolmogorov forward and backward equations or equivalently the Fokker-Planck equations associated to the forward and backward process:
\begin{align}
	\frac{\partial}{\partial \tau} \rho(x,\tau) 
	&= - \nabla_\mu \left[\hat{v}_+^\mu(x,\tau) \rho(x,\tau)\right] + \frac{\hbar}{2 m} \nabla^2 \rho(x,\tau),\nonumber\\
	\frac{\partial}{\partial \tau} \rho(x,\tau) 
	&= - \nabla_\mu \left[\hat{v}_-^\mu(x,\tau) \rho(x,\tau)\right] - \frac{\hbar}{2 m} \nabla^2 \rho(x,\tau).
\end{align}
Adding and subtracting the two equations leads to the continuity and osmotic equations
\begin{align}
	\frac{\partial}{\partial \tau} \rho(x,\tau) 
	&= - \nabla_\mu \left[v^\mu(x,\tau) \rho(x,\tau)\right],\label{eq:Continuity}\\
	\hat{u}^\mu(x,\tau)
	&= \frac{\hbar}{2 m} \nabla^\mu \ln\left[\rho(x,\tau)\right].\label{eq:Osmotic}
\end{align}

\section{The Stochastic Lagrangian}\label{sec:StochLag}
In classical physics a Lagrangian is a function of the form $L(X,V,\tau)$. In stochastic quantization on the other hand the Lagrangian is a function of the form $L(X,V_+,V_-,\tau)$. Due to the existence of two different velocities, it is not immediately clear how the classical Lagrangian should be generalized to the stochastic framework. However, it was shown by Zambrini, cf. Ref.~\cite{Zambrini} that for any classical Lagrangian of the form
\begin{equation}\label{eq:Clas}
	L_c(x,v,\tau) = \frac{m}{2} T_{\mu\nu}(x,\tau) v^\mu v^{\nu} - \hbar A_\mu(x,\tau) v^\mu - \mathfrak{U}(x,\tau)
\end{equation}
the minimal stochastic extension that is compatible with gauge invariance and Maupertuis' principle is given by
\begin{equation}
	L(X,V_+,V_-,\tau) = \frac{1}{2} L_c(X,V_+,\tau) + \frac{1}{2} L_c(X,V_-,\tau).
\end{equation}
We note that this form of the Lagrangian was also assumed by Yasue \cite{Yasue,Yasue:1981wu}. In the remainder of this paper, we will assume that gravity is the only spin-2 field, i.e.
\begin{equation}
	T_{\mu\nu}(x,\tau) = g_{\mu\nu}(x).
\end{equation}
The stochastic Lagrangian corresponding to the classical Lagrangian \eqref{eq:Clas} is then given by
\begin{equation}\label{eq:StochLagrangian}
	L\left(X,V_+,V_-\right) = \frac{m}{4} g_{\mu\nu} \left( V_+^\mu V_+^\nu + V_-^\mu V_-^\nu\right) - \frac{\hbar}{2} A_\mu(X)\left( V_+^\mu + V_-^\mu \right) - \mathfrak{U}(X)
\end{equation}
or equivalently
\begin{equation}\label{eq:StochLagrangian2}
	L\left(X,V,U\right) = \frac{m}{2} g_{\mu\nu} \left( V^\mu V^\nu + U^\mu U^\nu\right) - \hbar \, A_\mu(X) V^\mu  - \mathfrak{U}(X).
\end{equation}
Compared to the classical Lagrangian there is an additional energy contribution:
\begin{equation}
	\frac{m}{2} g_{\mu\nu} 
	U^\mu U^\nu.
\end{equation} 
This is the osmotic energy and can be interpreted as the kinetic energy of the background field.
\par

There also exists a Hamiltonian description. The momenta for this Lagrangian are
\begin{align}
	P^+_\mu(\tau) &= \frac{m}{2} g_{\mu\nu} V_+^\nu(\tau) - \frac{\hbar}{2} A_\mu(X),\nonumber\\
	P^-_\mu(\tau) &= \frac{m}{2} g_{\mu\nu} V_-^\nu(\tau) - \frac{\hbar}{2} A_\mu(X),\nonumber\\
	P_\mu(\tau) &= m \, g_{\mu\nu} V^\nu(\tau) - \hbar\, A_\mu(X),\nonumber\\
	Q_\mu(\tau) &= m \, g_{\mu\nu} U^\nu(\tau).
\end{align}
The Hamiltonian is then given by
\begin{equation}
	H\left(X,P^+,P^-\right) = \frac{1}{m} g^{\mu\nu} 
	\left(P^+_\mu P^+_\nu + P^-_\mu P^-_\nu + \hbar \left( P^+_\mu + P^-_\mu \right) A_\nu(X) + \frac{\hbar^2}{2} A_\mu(X) A_\nu(X) \right) 
	+ \mathfrak{U}(X)
\end{equation}
or equivalently
\begin{equation}
	H\left(X,P,Q \right) = \frac{1}{2m} g^{\mu\nu} 
	\left( P_\mu P_\nu + Q_\mu Q_\nu + 2 \hbar \, P_\mu A_\nu(X) + \hbar^2\, A_\mu(X) A_\nu(X) \right) 
	+ \mathfrak{U}(X).
\end{equation}

\subsection{Conditional expectations}\label{sec:CondExp}
In section \ref{sec:HamJac}, we derived the Hamilton-Jacobi equations and obtained expressions that contained the conditional expectation of the Lagrangian $\E_{\tau}\left[ L(X,V,U,\tau)\right]$. We can calculate this expression for the Lagrangian \eqref{eq:StochLagrangian} obtained in previous subsection. For this we notice that for any smooth function $\mathfrak{U}:T\times \M \rightarrow \R$
\begin{equation}
	\E_\tau\left[\mathfrak{U}(X(\tau),\tau)\right] = \lim_{s \rightarrow \tau} \E_\tau\left[\mathfrak{U}(X(s),s)\right] = \mathfrak{U}(X(\tau),\tau).
\end{equation}
For the terms that depend on the velocity process, we need to make sense of the processes $V_\pm$. This can be done by performing an integration over $d\tau$. At linear order we have
\begin{align}\label{eq:AVp}
	A_\mu(X(\tau),\tau) V_+^\mu(\tau) 
	&= \lim_{h\rightarrow 0} \frac{1}{h} \int_{\tau}^{\tau+h} A_\mu(X(s),s) V_+^\mu(s) \, ds  \nonumber\\
	&= \lim_{h\rightarrow 0} \frac{1}{h} \left[ \lowint_{\tau}^{\tau+h} \Big( 
	A_\mu(X(s),s) \, d_+X^\mu(s) + \p_\nu A_\mu(X(s),s) \, dX^\mu \cdot dX^\nu(s) \Big) \right]\nonumber\\
	&= \lim_{h\rightarrow 0} \frac{1}{h} \left[ \lowint_{\tau}^{\tau+h} \Big( 
	A_\mu(X(s),s) \, d_+\hat{X}^\mu(s) + \nabla_\nu A_\mu(X(s),s) \, d\hat{X}^\mu \cdot d\hat{X}^\nu(s) \Big) \right].
\end{align}
By a similar calculation, we obtain
\begin{align}\label{eq:AVm}
	A_\mu(X(\tau),\tau) V_-^\mu(\tau) 
	&= \lim_{h\rightarrow 0} \frac{1}{h} \left[ \upint_{\tau}^{\tau+h} \left( 
	A_\mu(X(s),s) \, d_-\hat{X}^\mu(s) - \nabla_\nu A_\mu(X(s),s) \, d\hat{X}^\mu \cdot d\hat{X}^\nu(s) \right) \right].
\end{align}
We note that we can write these expressions in differential notation as
\begin{equation}
	A_\mu V_\pm^\mu \, d\tau = A_\mu\, d_\pm \hat{X}^\mu \pm \nabla_\nu A_\mu \, d \hat{X}^\mu \cdot d \hat{X}^\nu
\end{equation}
Taking the expectation value of these expressions yields
\begin{align}
	\E_{\tau}\left[ A_\mu(X(\tau),\tau) V_+^\mu(\tau) \right] 
	&= 
	\lim_{h\rightarrow 0} \frac{1}{h} \E_{\tau} \left[ 
	\int_{\tau}^{\tau+h} A_\mu (X(s),s) \, \hat{v}_+^\mu (X(s),s) \, ds  \right. \nonumber\\
	&\qquad \qquad
	+ \lowint_{\tau}^{\tau+h} A_\mu (X(s),s) \, dW_+^\mu(s) \nonumber\\
	&\qquad \qquad \left.
	+ \int_{\tau}^{\tau+h} \nabla_\nu A_\mu(X(s),s)\, \hat{v}_+^{\mu\nu}(X(s),s) \, ds  \right]\nonumber\\
	&= A_\mu(X(\tau),\tau)\, \hat{v}_+^\mu(X,\tau) + \frac{\hbar}{2m} \nabla_\mu A^\mu(X(\tau),\tau),
\end{align}
where we used the martingale property \eqref{eq:MartingaleProperty}. Moreover,
\begin{align}
	\E_{\tau}\left[A_\mu(X(\tau),\tau)\, V_-^\mu(\tau) \right]
	&=
	A_\mu(X(\tau),\tau)\, \hat{v}_-^\mu(X,\tau) - \frac{\hbar}{2m} \nabla_\mu A^\mu(X(\tau),\tau).
\end{align}
Consequently,
\begin{align}
	\E_{\tau}\left[A_\mu(X(\tau),\tau)\, V^\mu(\tau) \right]
	&=
	A_\mu(X(\tau),\tau)\, v^\mu(X,\tau),\\
	\E_{\tau}\left[A_\mu(X(\tau),\tau)\, U^\mu(\tau) \right]
	&=
	A_\mu(X(\tau),\tau)\, \hat{u}^\mu(X,\tau) + \frac{\hbar}{2m} \nabla_\mu A^\mu(X(\tau),\tau).
\end{align}
\par

For the terms quadratic in velocity we will perform a double integral over $d\tau$. In differential notation we have\footnote{cf. section 9 in Ref.~\cite{Nelson}.}
\begin{align}
	g_{\mu\nu} V_+^\mu V_+^\nu \, d\tau^2
	&=
	g_{\mu\nu} \, d_+ \hat{X}^\mu \otimes d_+ \hat{X}^\nu 
	+ g_{\mu\nu} \, \nabla_\rho \left( d_+ \hat{X}^\mu \right) \otimes d\hat{X}^\nu \cdot d\hat{X}^\rho\nonumber\\
	&\quad
	+ g_{\mu\nu} \,  d\hat{X}^\mu \cdot d\hat{X}^\rho \otimes \nabla_\rho \left( d_+ \hat{X}^\nu \right)
	- \frac{2}{3} \mathcal{R}_{\mu\nu\rho\sigma} \, d \hat{X}^\mu \cdot d \hat{X}^\rho \otimes d \hat{X}^\nu \cdot d \hat{X}^\sigma,\nonumber\\
	g_{\mu\nu} V_-^\mu V_-^\nu \, d\tau^2
	&=
	g_{\mu\nu} \, d_- \hat{X}^\mu \otimes d_- \hat{X}^\nu 
	- g_{\mu\nu} \, \nabla_\rho \left( d_- \hat{X}^\mu \right) \otimes d\hat{X}^\nu \cdot d\hat{X}^\rho\nonumber\\
	&\quad
	- g_{\mu\nu} \,  d\hat{X}^\mu \cdot d\hat{X}^\rho \otimes \nabla_\rho \left( d_- \hat{X}^\nu \right)
	- \frac{2}{3} \mathcal{R}_{\mu\nu\rho\sigma} \, d \hat{X}^\mu \cdot d \hat{X}^\rho \otimes d \hat{X}^\nu \cdot d \hat{X}^\sigma,\nonumber\\
	g_{\mu\nu} V_+^\mu V_-^\nu \, d\tau^2
	&=
	g_{\mu\nu} \, d_+ \hat{X}^\mu \otimes d_- \hat{X}^\nu 
	- g_{\mu\nu} \, \nabla_\rho \left( d_+ \hat{X}^\mu \right) \otimes d\hat{X}^\nu \cdot d\hat{X}^\rho\nonumber\\
	&\quad
	+ g_{\mu\nu} \, d\hat{X}^\mu \cdot d\hat{X}^\rho \otimes \nabla_\rho \left( d_- \hat{X}^\nu \right) 
	+ \frac{2}{3} \mathcal{R}_{\mu\nu\rho\sigma} \, d \hat{X}^\mu \cdot d \hat{X}^\rho \otimes d \hat{X}^\nu \cdot d \hat{X}^\sigma.\label{eq:QuadVelProc} 
\end{align}
We can take the expectation values of these expressions. This yields 
\begin{align}
	\E_{\tau}\Big[ g_{\mu\nu} \, d_+ \hat{X}^\mu \otimes d_+ \hat{X}^\nu \Big]
	&=
	\E_{\tau}\Big[ g_{\mu\nu} \Big(
	\hat{v}_+^\mu \hat{v}_+^\nu \, d\tau^2
	+ \hat{v}_+^\mu \, dW_+^\nu \, d\tau
	+ \hat{v}_+^\nu \, dW_+^\nu \, d\tau
	+ dW_+^\mu \otimes dW_+^\nu \Big) \Big]\nonumber\\
	&=
	g_{\mu\nu} \Big(
	\hat{v}_+^\mu \hat{v}_+^\nu \, d\tau^2
	+ 2 \, \hat{v}_+^{\mu\nu} \, d\tau  \Big)\nonumber\\
	&=
	\frac{n\, \hbar}{m} \, d\tau
	+ g_{\mu\nu} \hat{v}_+^\mu \hat{v}_+^\nu \, d\tau^2
\end{align}
where we used that the expectation value of the terms linear in $dW_+$ vanishes, due to the martingale property of $W_+$. Moreover, we used eq.~\eqref{eq:IntBiForm} to evaluate the term $dW_+^\mu dW_+^\nu = dW_+^\mu \otimes dW_+^\nu$. By a similar calculation we obtain
\begin{align}
	\E_{\tau}\Big[ g_{\mu\nu} \; d_- \hat{X}^\mu \otimes d_- \hat{X}^\nu \Big]
	&=
	- \frac{n\,\hbar}{m} \, d\tau
	+ g_{\mu\nu} \hat{v}_-^\mu \hat{v}_-^\nu \, d\tau^2\\
	\E_{\tau}\Big[ g_{\mu\nu} \; d_+ \hat{X}^\mu \otimes d_- \hat{X}^\nu \Big]
	&=
	g_{\mu\nu} \hat{v}_+^\mu \hat{v}_-^\nu \, d\tau^2.
\end{align}
Furthermore,
\begin{align}
	\E_{\tau}\Big[ g_{\mu\nu} \nabla_{\rho} \left( d_+ \hat{X}^\mu \right) \otimes d\hat{X}^\nu \cdot d\hat{X}^\rho \Big]
	&=
	\E_{\tau}\Big[ g_{\mu\nu} \hat{v}_+^{\nu\rho} \nabla_{\rho} \left(  \hat{v}_+^\mu d\tau + dW_+^\mu \right) d\tau \Big]\nonumber\\
	&=
	\frac{\hbar}{2m} \nabla_\mu \hat{v}_+^\mu \, d\tau^2.
\end{align}
Similarly,
\begin{align}
	\E_{\tau}\Big[ g_{\mu\nu} \nabla_{\rho} \left( d_- \hat{X}^\mu \right) \otimes d\hat{X}^\nu \cdot d\hat{X}^\rho \Big]
	&=
	\frac{\hbar}{2m} \nabla_\mu \hat{v}_-^\mu \, d\tau^2.
\end{align}
For the remaining term we find
\begin{align}
	\E_{\tau}\left[ \mathcal{R}_{\mu\nu\rho\sigma} \, d \hat{X}^\mu \cdot d \hat{X}^\rho \otimes d \hat{X}^\nu \cdot d \hat{X}^\sigma \right]
	&=
	\E_{\tau}\left[ \mathcal{R}_{\mu\nu\rho\sigma} \, \hat{v}^{\mu\rho} \hat{v}^{\nu\sigma} d\tau^2 \right] \nonumber\\
	&=
	\frac{\hbar^2}{4m^2} \mathcal{R} \, d\tau^2.
\end{align}
We conclude,
\begin{align}
	\E_\tau \left[g_{\mu\nu} \, V_+^\mu V_+^\nu \right] 
	&= 
	g_{\mu\nu} \hat{v}_+^\mu \hat{v}_+^\nu 
	+ \frac{\hbar}{m} \nabla_\mu \hat{v}_+^\mu
	- \frac{\hbar^2}{6 m^2}  \mathcal{R}
	+ \frac{n\, \hbar}{m \, d\tau},\nonumber\\
	\E_\tau \left[g_{\mu\nu} \, V_-^\mu V_-^\nu \right] 
	&= 
	g_{\mu\nu} \hat{v}_-^\mu \hat{v}_-^\nu 
	- \frac{\hbar}{m} \nabla_\mu \hat{v}_-^\mu
	- \frac{\hbar^2}{6 m^2} \mathcal{R}
	- \frac{n\, \hbar}{m \, d\tau},\nonumber\\
	\E_\tau\left[g_{\mu\nu}  \, V_+^\mu V_-^\nu \right] 
	&= 
	g_{\mu\nu} \hat{v}_+^\mu \hat{v}_-^\nu
	- \frac{\hbar}{2m} \nabla_\mu \hat{v}_+^\mu
	+ \frac{\hbar}{2m} \nabla_\mu \hat{v}_-^\mu
	+ \frac{\hbar^2}{6 m^2} \mathcal{R}
\end{align}
or equivalently
\begin{align}
	\E_\tau \Big[g_{\mu\nu} \, V^\mu V^\nu \Big] 
	&= 
	g_{\mu\nu} v^\mu v^\nu,\nonumber\\
	\E_\tau \left[g_{\mu\nu} \, U^\mu U^\nu \right] 
	&= 
	g_{\mu\nu} \hat{u}^\mu  \hat{u}^\nu 
	+ \frac{\hbar}{m} \nabla_\mu \hat{u}^\mu
	- \frac{\hbar^2}{6 m^2} \mathcal{R},\nonumber\\
	\E_\tau \left[g_{\mu\nu} \, V^\mu U^\nu \right] 
	&= 
	g_{\mu\nu} v^\mu \hat{u}^\nu + \frac{\hbar}{2m}\nabla_\mu  v^\mu
	+ \frac{n\, \hbar}{2\,m \, d\tau}.
\end{align}
The conditional expectation of the Lagrangian \eqref{eq:StochLagrangian2} is thus given by\footnote{Note that the divergent term $ \frac{n\hbar}{m  d\tau}$ does not appear in the Lagrangian.}
\begin{align}\label{eq:StochLagrangianExpectation}
	\E_\tau \left[L(X,V,U,\tau)\right] 
	&= 
	\frac{m}{2} g_{\mu\nu} \left(  v^\mu v^\nu + \hat{u}^\mu \hat{u}^\nu \right)
	+ \frac{\hbar}{2} \nabla_\mu \hat{u}^\mu 
	- \frac{\hbar^2}{12 m} \mathcal{R}
	- \hbar \, A_\mu v^\mu
	- \mathfrak{U}.
\end{align}

\subsection{Correlation functions}
Observables in quantum mechanics can be constructed from correlation functions computed in the path integral formalism. Since this computation is slightly different in stochastic quantization, we review the main steps. 
\par 

In order to compute correlation functions within the stochastic quantization, one must first solve the stochastic equations of motion derived from the action. The solution is a stochastic process $\{X(\tau)|\tau\in T\}$. For this stochastic process one can define a characteristic functional $\Phi_{X}(J)$, and a moment generating functional $M_{X}(J)$:
\begin{align}
	\Phi_{X}(J) 
	&= \E\left[ e^{\frac{i}{\hbar} \int_{\tau_i}^{\tau_f} J_\mu(\tau) X^\mu(\tau) d\tau }\right],\\
	M_{X}(J)
	&= \E\left[ e^{\frac{1}{\hbar} \int_{\tau_i}^{\tau_f} J_\mu(\tau) X^\mu(\tau) d\tau } \right],
\end{align}
where $J(\tau)$ is a bounded process of finite variation that corresponds to the source in the path integral formulation. We emphasize that one no longer averages over the action, as this is essentially done in the first step, where the stochastic differential equation is solved. 
\par 

Using the characteristic and moment generating functionals for the process $X(\tau)$, one can calculate all moments of the theory. For example, the two-point correlation function is given by
\begin{equation}
	\E\left[ X^\mu(s) X^\nu(r) \right] 
	= \lim_{||J||\rightarrow 0} \frac{\p}{\p J_\mu(s)} \frac{\p}{\p J_\nu(r)} M_{X}(J).
\end{equation}
\par

We emphasize that the integrals that need to be evaluated in the path integral formalism and stochastic quantization are constructed in different ways. Due to this different construction, theories that require renormalization in the path integral formalism can be finite in stochastic quantization.

\subsection{Uncertainty principle}\label{sec:Uncertainty}
Due to the relevance of the unvertainty principle in quantum mechanics, we will derive it in stochastic quantization, which can be done using the results from section~\ref{sec:CondExp}.
\par 

For $s>\tau$ we find
\begin{align}
	{\rm Cov}_\tau \left[X_\mu(s), X^\nu(s) \right]
	&=
	\E_{\tau}\left[ X_\mu(s) X^\nu(s) \right] 
	- \E_{\tau}\left[ X_\mu(s) \right] \E_{\tau}\left[ X^\nu(s) \right]\nonumber\\
	&=
	\E_{\tau}\left[ \left( X_\mu(\tau) + \int_{\tau}^{s} V_{+\mu}(r)\, dr \right) \left( X^\mu(\tau) + \int_{\tau}^{s} V_+^\mu(r)\, dr \right) \right] \nonumber\\
	&\quad
	- \E_{\tau}\left[ \left( X_\mu(\tau) + \int_{\tau}^{s} V_\mu(r) \, dr \right) \right] 
	\E_{\tau}\left[ \left( X^\mu(\tau) + \int_{\tau}^{s} V^\mu(r) \, dr \right) \right]  \nonumber\\
	&=
	\frac{\hbar}{m} \delta_\mu^\nu (s-\tau)
	+ \frac{\hbar}{2m} \left(\nabla_\mu \hat{v}_+^\nu + \nabla^\nu \hat{v}_{+\mu} - \frac{\hbar}{3m} \mathcal{R}_\mu^\nu \right) (s-\tau)^2
	+ o(s-\tau)^2.
\end{align}
Furthermore, the covariance for the momenta is given by
\begin{align}
	{\rm Cov}_\tau \left[P^+_\mu(s), P^{+\nu}(s) \right]
	&=
	\frac{m^2}{4} \Big\{
	\E_{\tau}\left[V_{+\mu}(s) V_+^\nu(s) \right]
	- \E_{\tau}\left[ V_{+\mu}(s)\right] \E_{\tau}\left[V_+^\nu(s) \right] \Big\}\nonumber\\
	&\quad
	- \frac{m\,\hbar}{4}\Big\{
	\E_{\tau}\left[ V_{+\mu}(s) A^\nu(s) \right]
	- \E_{\tau}\left[ V_{+\mu}(s)\right] \E_{\tau}\left[ A^\nu(s) \right]\Big\}\nonumber\\
	&\quad
	- \frac{m\,\hbar}{4}\Big\{
	\E_{\tau}\left[ A_\mu(s) V_+^\nu(s) \right]
	- \E_{\tau}\left[ A_\mu(s) \right] \E_{\tau}\left[ V_+^\nu(s) \right] \Big\}\nonumber\\
	&\quad
	- \frac{\hbar^2}{4}\Big\{
	\E_{\tau}\left[ A_\mu(s) A^\nu(s) \right]
	- \E_{\tau}\left[ A_\mu(s) \right] \E_{\tau}\left[ A^\nu(s) \right] \Big\}\nonumber\\
	&=
	\frac{m\,\hbar}{4} \delta_\mu^\nu (s-\tau)^{-1}
	+ \frac{m\,\hbar}{8} \left(\nabla_\mu \hat{v}_+^\nu + \nabla^\nu \hat{v}_{+\mu}\right) \nonumber\\
	&\quad
	- \frac{\hbar^2}{8} \left(\nabla_\mu A^\nu + \nabla^\nu A_\mu \right)
	- \frac{\hbar^2}{24} \mathcal{R}_\mu^\nu
	+ o(1).
\end{align}
If we take the limit $s\rightarrow \tau$, we find
\begin{align}
	\lim_{s\rightarrow \tau} {\rm Cov}_\tau \left[X_\mu(s), X^\nu(s) \right]
	&= 0, \\
	\lim_{s\rightarrow \tau} {\rm Cov}_\tau \left[P^+_\mu(s), P^{+\nu}(s) \right]
	&= \infty.
\end{align}
This reflects the fact that we have constructed the stochastic theory in a position representation, i.e. the process $(X,P_+,P_-)$ is adapted to the filtration generated by the process $X$.
\par 

We can calculate the product of the two variances. For this we fix the indices $\mu=\nu=\bar{\mu}$, and obtain
\begin{align}
	{\rm Var}_\tau \left[X^{\bar{\mu}} (s) \right]
	{\rm Var}_\tau \left[P^+_{\bar{\mu}} (s) \right]
	&=
	\frac{\hbar^2}{4}
	+ \frac{\hbar^2}{2} \left( \nabla_{\bar{\mu}} \hat{v}_+^{\bar{\mu}}
	- \frac{\hbar}{2m} \, \nabla_{\bar{\mu}} A^{\bar{\mu}} 
	- \frac{\hbar}{6m} \mathcal{R}_{\bar{\mu}}^{\bar{\mu}} \right)  (s-\tau)
	+ o(s-\tau).
\end{align}
If we then take the limit $s\rightarrow\tau$, we find
\begin{equation}
	\lim_{s\rightarrow \tau} {\rm Var}_\tau \left[X^{\bar{\mu}} (s) \right]
	{\rm Var}_\tau \left[P^+_{\bar{\mu}} (s) \right]
	=
	\frac{\hbar^2}{4}.
\end{equation}
This corresponds to the lower bound given by the Heisenberg uncertainty principle.

\section{Scalar Test Particles}
In this section, we derive the equations of motion that govern a quantum mechanical spin-0 test particle on a pseudo-Riemannian manifold subjected to the Lagrangian \eqref{eq:StochLagrangian2}. 

\subsection{Stochastic equation of motion}\label{sec:StochEQM}
We consider the Lagrangian \eqref{eq:StochLagrangian2}:
\begin{align}
	L\left(X,V,U\right) 
	&= 
	\frac{m}{2} g_{\mu\nu} \left( V^\mu V^\nu + U^\mu U^\nu\right) 
	- \hbar \, A_\mu V^\mu  - \mathfrak{U}.
\end{align}
After integrating this expression twice over $\tau$ we obtain, cf. eq.~\eqref{eq:QuadVelProc},
\begin{align}
	\mathbb{E} \left[ L\, d\tau^2 \right]
	&=
	\mathbb{E} \left[ \frac{m}{2}   g_{\mu\nu} \left\{ 
	 dX^\mu dX^\nu + d_\circ\hat{X}^\mu d_\circ\hat{X}^\nu 
	+ \nabla_\rho \left( d_\circ \hat{X}^\mu \right) d[[X^\nu,X^\rho]] \right. \right. \nonumber\\
	&\qquad \qquad \qquad \left. 
	- \frac{\hbar^2}{6 m^2} \mathcal{R}^\mu_{\;\;\rho\kappa\sigma} \, d[[X^\nu,X^\kappa]] \, d[[X^\rho,X^\sigma]]
	\right\} \nonumber\\
	&\qquad \quad \left.
	- \hbar \, A_\mu \, dX^\mu \, d\tau
	-  \mathfrak{U} \, d\tau^2 \right] \nonumber\\
	&=
	\mathbb{E} \left[ \frac{m}{2}  g_{\mu\nu} \, dX^\mu dX^\nu 
	- \hbar \, A_\mu dX^\mu d\tau
	-  \left( \mathfrak{U} + \frac{\hbar^2}{12m} \mathcal{R} \right) d\tau^2 \right],
\end{align}
where we used
\begin{equation}
	\E \left[ g_{\mu\nu} \left\{ d_\circ \hat{X}^\mu d_\circ \hat{X}^\nu 
	+ \nabla_\rho \left( d_\circ \hat{X}^\mu \right) d[[X^\nu,X^\rho]] \right\} \right]
	=
	0,
\end{equation}
which follows from eq.~\eqref{eq:OsmoticIntegral2} and the metric compatibility. If we vary this expression with respect to a stochastically independent deviation process $\delta X$, we obtain the stochastic Euler-Lagrange equations \eqref{eq:StochEL3} that take the form
\begin{align}\label{eq:EQM}
	m \, \left( 
	g_{\mu\nu} d^2 X^\nu 
	+ g_{\mu\nu} \Gamma^\nu_{\rho\sigma} dX^\rho dX^\sigma 
	\right)
	&=
	\left(
	\hbar\, \p_\tau A_\mu - \nabla_\mu \mathfrak{U} - \frac{\hbar^2}{12 m} \nabla_\mu \mathcal{R}  
	\right) d\tau^2
	- \hbar\, H_{\mu\nu}\, dX^\nu d\tau,
\end{align}
where
\begin{equation}
	H_{\mu\nu} := \p_\mu A_\nu - \p_\nu A_\mu
	=\nabla_\mu A_\nu - \nabla_\nu A_\mu.
\end{equation}
\par 

In the classical limit $\hbar\rightarrow0$, the quadratic variation vanishes. This gives\footnote{Note that $\mathfrak{U}$ and $A_\mu$ could contain an additional $\hbar$ dependence.}
\begin{equation}
	m \left( 
	g_{\mu\nu} \frac{d^2 X^\nu}{d\tau^2} 
	+ g_{\mu\nu} \Gamma^\nu_{\rho\sigma} \frac{dX^\rho}{d\tau} \frac{dX^\sigma}{d\tau} 
	\right)
	=
	- \lim_{\hbar\rightarrow 0} \left\{ 
	\nabla_\mu \mathfrak{U}
	+ \hbar \left[
	-\p_\tau A_\mu
	+ H_{\mu\nu} \frac{dX^\nu}{d\tau} \right] \right\},
\end{equation}
which is consistent with general relativity. On the other hand, taking the flat space-time limit $G_{\rm N}\rightarrow0$ gives $g_{\mu\nu} = \eta_{\mu\nu}$, and therefore
\begin{equation}
	m \, 
	\eta_{\mu\nu} \, d^2 X^\nu
	=
	\big(
	\hbar\, \p_\tau A_\mu - \p_\mu \mathfrak{U}  
	\big)\, d\tau^2
	- \hbar \, H_{\mu\nu}\, dX^\nu d\tau.
\end{equation}
If we then take the non-relativistic limit $c\rightarrow\infty$, we identify $t=\tau$ and replace $\eta_{\mu\nu}\rightarrow\delta_{ij}$. The resulting equation is consistent with stochastic quantization in flat spaces \cite{Kershaw,Nelson:1966sp,NelsonOld,Nelson,Zambrini,Guerra:1981ie}.
\par 

The stochastic differential equation \eqref{eq:EQM} is the fundamental equation of motion in stochastic quantization. The solutions describe the stochastic trajectories of quantum mechanical spin-0 test particles in any geometry. In section \ref{sec:Schrodinger}, we will show that probability density function associated to the solution $X(\tau)$ of this equation evolves according to the Schr\"odinger equation.
\subsection{Stochastic Newton equation}\label{sec:StochNewton}
The stochastic differential equation derived in previous section can be rewritten as a diffusion equation for the vector fields $v_\pm(x,\tau)$. This representation is known as the stochastic Newton equation, see e.g. Ref.~\cite{Nelson}. In order to derive it, we define a function
\begin{equation}\label{eq:Rdef}
	R(x,\tau) :=\frac{\hbar}{2} \ln\left[\rho(x,\tau)\right].
\end{equation}
The osmotic \eqref{eq:Osmotic} and continuity equation \eqref{eq:Continuity} can then be rewritten as
\begin{align}
	\nabla^\mu R(x,\tau) &= m\, \hat{u}^\mu,\label{eq:Osmotic2}\\
	\frac{\p}{\p \tau} R(x,\tau) &= - \left(m\,g_{\mu\nu} \hat{u}^\nu + \frac{\hbar}{2} \nabla_\mu \right)\hat{v}^\mu.\label{eq:Continuity2}
\end{align}
Furthermore, we recall that the Hamilton Jacobi equations \eqref{eq:HamJac1} and \eqref{eq:HamJac2} are given by
\begin{align}
	\nabla_{\mu} S(x,\tau) &= p_\mu,\\
	\frac{\p}{\p \tau} S(x,\tau) &= E_\tau\left[L(X,V,U,\tau)\right] - p_\mu v^\mu.
\end{align}
We consider the Lagrangian \eqref{eq:StochLagrangian2}
\begin{equation}
	L(X,V,U,\tau) 
	= 
	\frac{m}{2} g_{\mu\nu} \left( V^\mu V^\mu + U^\mu U^\mu \right) - \hbar \, A_\mu V^\mu - \mathfrak{U}
\end{equation}
with momenta
\begin{align}
	P_{\mu}(\tau) 
	&= 
	m\,g_{\mu\nu} V^\nu - \hbar \, A_\mu,\nonumber\\
	Q_{\mu}(\tau) 
	&= 
	m\, g_{\mu\nu} U^\nu.	
\end{align}
Therefore,
\begin{align}
	p_{\mu}(x,\tau) 
	&=
	\E_\tau\left[ P_\mu(\tau)\right] 
	= 
	m\, g_{\mu\nu} v^\nu - \hbar \, A_\mu,\nonumber\\
	\hat{q}_{\mu}(x,\tau) 
	&= 
	\E_\tau\left[ Q_\mu(\tau)\right] 
	= 
	m\, g_{\mu\nu} \hat{u}^\nu.	
\end{align}
Moreover, in eq.~\eqref{eq:StochLagrangianExpectation}, we found
\begin{equation}
	\E_\tau \left[L(X,V,U,\tau)\right] 
	= 
	\frac{m}{2} g_{\mu\nu} \left(  v^\mu v^\nu + \hat{u}^\mu \hat{u}^\nu \right)
	+ \frac{\hbar}{2} \nabla_\mu \hat{u}^\mu 
	- \frac{\hbar^2}{12 m} \mathcal{R} 
	- \hbar \, A_\mu v^\mu
	- \mathfrak{U}.
\end{equation}
Putting everything together yields
\begin{align}
	\nabla_{\mu} S(x,\tau) 
	&= 
	p_\mu 
	= 
	m\, g_{\mu\nu} v^\nu - \hbar \, A_\mu, \label{eq:NablaS}\\
	\nabla_{\mu} R(x,\tau) 
	&= 
	\hat{q}_\mu  
	= 
	m\, g_{\mu\nu} \hat{u}^\nu \label{eq:NablaR}
\end{align}
and
\begin{align}
	\frac{\p}{\p \tau} S(x,\tau) 
	&= 
	- \frac{m}{2} g_{\mu\nu} \left(v^\mu v^\nu
	- \hat{u}^\mu \hat{u}^\nu\right)
	+ \frac{\hbar}{2} \nabla_\mu \hat{u}^\mu  
	- \frac{\hbar^2}{12 m} \mathcal{R}
	- \mathfrak{U}, \label{eq:HamJac23}\\
	\frac{\p}{\p \tau} R(x,\tau) \label{}
	&= 
	- m\, g_{\mu\nu} v^\mu \hat{u}^\mu
	- \frac{\hbar}{2}\nabla_\mu v^\mu.\label{eq:Continuity3}
\end{align}
\par 

We take the covariant derivative of eq.~\eqref{eq:HamJac23}. This yields
\begin{equation}
	m \frac{\p v_\mu}{\p \tau} 
	- \hbar \frac{\p A_\mu}{\p \tau}
	= 
	- m \, v^\nu \nabla_\mu v_\nu
	+ m \, \hat{u}^\nu \nabla_\mu \hat{u}_\nu 
	+ \frac{\hbar}{2} \nabla_\mu \nabla_\nu \hat{u}^\nu  
	- \frac{\hbar^2}{12 m} \nabla_\mu \mathcal{R} 
	- \nabla_\mu \mathfrak{U}.
\end{equation}
Using eqs.~\eqref{eq:NablaS} and \eqref{eq:NablaR}, we find
\begin{align}
	\nabla_\mu \hat{u}_\nu &= \nabla_\nu \hat{u}_\mu,\nonumber\\
	 \nabla_\mu v_\nu &= \nabla_\nu v_\mu + \frac{\hbar}{m} H_{\mu\nu},\nonumber\\
	 \nabla_\mu \nabla_\nu \hat{u}^\nu &= \Box \hat{u}_\mu - \mathcal{R}_{\mu\nu} \hat{u}^\nu.
\end{align}
Therefore,
\begin{align}
	\hbar\left( \frac{\p A_\mu}{\p \tau} - H_{\mu\nu} \hat{v}^\nu \right)
	- \frac{\hbar^2}{12 m} \nabla_\mu \mathcal{R} 
	- \nabla_\mu \mathfrak{U}
	&=
	m \left(\frac{\p \hat{v}_\mu}{\p \tau} 
	+ v^\nu \nabla_\nu \hat{v}_\mu 
	- \hat{u}^\nu \nabla_\nu \hat{u}_\mu \right. \nonumber\\
	& \qquad \qquad 
	- \hat{u}^{\rho\sigma} \nabla_\rho \nabla_\sigma \hat{u}_\mu
	+ \hat{u}^{\rho\sigma} \mathcal{R}^\nu_{\;\;\rho\sigma\mu} \hat{u}_\nu \Big) .
\end{align}
We will associate the left hand side with a force, i.e. 
\begin{equation}
	F_\mu := \hbar\left( \frac{\p A_\mu}{\p \tau} - H_{\mu\nu} \hat{v}^\nu \right)
	- \frac{\hbar^2}{12 m} \nabla_\mu \mathcal{R} 
	- \nabla_\mu \mathfrak{U}.
\end{equation}
Moreover, we rewrite the left hand side in terms of the forward and backward velocity. We find
\begin{align}\label{eq:SNewton1}
	F^\mu
	&=
	\frac{m}{2} \left[ \left(
	\frac{\p}{\p \tau} + \hat{v}_+^\nu \nabla_\nu + \hat{v}_+^{\rho\sigma} \nabla_\rho \nabla_\sigma \right) \hat{v}_-^\mu -  \mathcal{R}^\mu_{\;\;\rho\sigma\nu} \hat{v}_+^{\rho\sigma} \hat{v}_-^\nu  \right. \nonumber\\
	& \qquad \quad \left.
	+ \left(\frac{\p}{\p \tau} + \hat{v}_-^\nu \nabla_\nu + \hat{v}_-^{\nu\rho} \nabla_\nu \nabla_\rho \right) \hat{v}_+^\mu - \mathcal{R}^\mu_{\;\;\rho\sigma\nu} \hat{v}_-^{\rho\sigma} \hat{v}_+^\nu \right].
\end{align}

As we would like to associate the right hand side with an acceleration, we define second order acceleration vectors $a_{\pm\pm}$ by 
\begin{align}
	a_{+\pm}^{\mu}(x,\tau) 
	&:= \lim_{h\rightarrow 0} \frac{1}{h} \E_\tau \big[V_\pm^\mu(\tau+h) - V_\pm^\mu(\tau) \big],\nonumber\\
	a_{-\pm}^{\mu}(x,\tau) 
	&:= \lim_{h\rightarrow 0} \frac{1}{h} \E_\tau \big[V_\pm^\mu(\tau) - V_\pm^\mu(\tau-h) \big],
\end{align}
and
\begin{align}
	a_{+\pm}^{\rho\sigma}(x,\tau) 
	&:= \lim_{h\rightarrow 0} 
	\frac{1}{2h} \E_\tau \Big\{ 
	\big[V_\pm^\rho(\tau+h) - V_\pm^\rho(\tau) \big] 
	\big[X^\sigma(\tau+h) - X^\sigma(\tau) \big] \Big\}\nonumber\\
	& \qquad \quad
	+ \frac{1}{2h} \E_\tau \Big\{ 
	\big[X^\rho(\tau+h) - X^\rho(\tau) \big] 
	\big[V_\pm^\sigma(\tau+h) - V_\pm^\sigma(\tau) \big] \Big\},\nonumber\\
	a_{-\pm}^{\rho\sigma}(x,\tau) 
	&:= \lim_{h\rightarrow 0} \frac{1}{2h} \E_\tau \Big\{
	\big[V_\pm^\rho(\tau) - V_\pm^\rho(\tau-h) \big]
	\big[X^\sigma(\tau) - X^\sigma(\tau-h) \big] \Big\}\nonumber\\
	& \qquad \quad
	+ \frac{1}{2h} \E_\tau \Big\{
	\big[X^\rho(\tau) - X^\rho(\tau-h) \big]
	\big[V_\pm^\sigma(\tau) - V _\pm^\sigma(\tau-h) \big] \Big\}.
\end{align}
Using the parallel transport equation \eqref{eq:DirStochCovDev}, we then find 
\begin{align}
	a_{+\pm}^\mu 
	&=
	\lim_{d\tau\rightarrow 0} \frac{1}{d\tau} \E_{\tau} \Big[d_+ \hat{v}_\pm^\mu + \Gamma^\mu_{\nu\rho} \hat{v}_\pm^\nu d_+x^\rho
	+ \left( \p_\nu \Gamma^\mu_{\rho\sigma} + \Gamma^\mu_{\nu\kappa} \Gamma^\kappa_{\rho\sigma} - 2 \Gamma^\mu_{\rho\kappa} \Gamma^\kappa_{\nu\sigma} \right) \hat{v}_\pm^\nu dx^\rho \cdot dx^\sigma 
	+ o(d\tau)\Big]\nonumber\\
	&=
	\p_\tau \hat{v}_\pm^\mu
	+ v_+^\nu \p_\nu \hat{v}_\pm^\mu 
	+ v_+^{\rho\sigma} \p_\rho \p_\sigma \hat{v}_\pm^\mu 
	+ \Gamma^\mu_{\nu\rho} \hat{v}_\pm^\nu v_+^\rho 
	+ \left( \p_\nu \Gamma^\mu_{\rho\sigma} + \Gamma^\mu_{\nu\kappa} \Gamma^\kappa_{\rho\sigma} - 2 \Gamma^\mu_{\rho\kappa} \Gamma^\kappa_{\nu\sigma} \right) \hat{v}_\pm^\nu v_+^{\rho\sigma} \nonumber\\
	&=
	\p_\tau \hat{v}_\pm^\mu
	+ \hat{v}_+^\nu \nabla_\nu \hat{v}_\pm^\mu 
	+ \hat{v}_+^{\rho\sigma} \nabla_\rho \nabla_\sigma \hat{v}_\pm^\mu 
	- 2 \Gamma^\mu_{\nu\rho} \hat{v}_+^{\rho\sigma} \nabla_\sigma \hat{v}_\pm^\nu 
	- \mathcal{R}^\mu_{\;\;\rho\sigma\nu} \hat{v}_+^{\rho\sigma} \hat{v}_\pm^\nu \label{eq:AccStoch1p}
\end{align}
and
\begin{align}
	a_{-\pm}^\mu
	&=
	\lim_{d\tau\rightarrow 0} \frac{1}{d\tau} \E_{\tau} \Big[d_- \hat{v}_\pm^\mu + \Gamma^\mu_{\nu\rho} \hat{v}_\pm^\nu d_-x^\rho
	- \left( \p_\nu \Gamma^\mu_{\rho\sigma} + \Gamma^\mu_{\nu\kappa} \Gamma^\kappa_{\rho\sigma} - 2 \Gamma^\mu_{\rho\kappa} \Gamma^\kappa_{\nu\sigma} \right) \hat{v}_\pm^\nu dx^\rho \cdot dx^\sigma
	+ o(d\tau) \Big]\nonumber\\
	&=
	\p_\tau \hat{v}_\pm^\mu
	+ \hat{v}_-^\nu \nabla_\nu \hat{v}_\pm^\mu 
	+ \hat{v}_-^{\rho\sigma} \nabla_\rho \nabla_\sigma \hat{v}_\pm^\mu 
	- 2 \Gamma^\mu_{\nu\rho} \hat{v}_-^{\rho\sigma} \nabla_\sigma \hat{v}_\pm^\nu 
	- \mathcal{R}^\mu_{\;\;\rho\sigma\nu} \hat{v}_-^{\rho\sigma} \hat{v}_\pm^\nu,\label{eq:AccStoch1m}
\end{align}
where we allow for an explicit proper-time dependence of the velocity $v_\pm(X,\tau)$. For the second order parts we find
\begin{align}
	a_{+\pm}^{\rho\sigma}
	&=
	\lim_{d\tau\rightarrow 0} \frac{2}{d\tau} \E_\tau\left[d\hat{v}_\pm^{(\rho} \cdot dx^{\sigma)} + \Gamma^{(\rho|}_{\kappa\lambda} \hat{v}_\pm^\kappa\, d x^\lambda \cdot d x^{|\sigma)} + o(d\tau)\right]\nonumber\\
	&=
	\hat{v}_+^{\rho\kappa} \nabla_\kappa \hat{v}_\pm^\sigma
	+ \hat{v}_+^{\kappa\sigma} \nabla_\kappa \hat{v}_\pm^\rho \label{eq:AccStoch2p}
\end{align}
and
\begin{align}
	a_{-\pm}^{\rho\sigma}
	&=
	\lim_{d\tau\rightarrow 0} \frac{2}{d\tau} \E_\tau\left[- d\hat{v}_\pm^{(\rho} \cdot dx^{\sigma)} - \Gamma^{(\rho|}_{\kappa\lambda} \hat{v}_\pm^\kappa\, d x^\lambda \cdot d x^{|\sigma)} + o(d\tau)\right]\nonumber\\
	&=
	\hat{v}_-^{\rho\kappa} \nabla_\kappa \hat{v}_\pm^\sigma
	+ \hat{v}_-^{\kappa\sigma} \nabla_\kappa \hat{v}_\pm^\rho.\label{eq:AccStoch2m}
\end{align}
Eq.~\eqref{eq:SNewton1} can now be rewritten as the stochastic Newton equation
\begin{equation}
	F^\mu(X,\tau) =\frac{1}{2} m\left[ \hat{a}_{+-}^\mu(X,\tau) + \hat{a}_{-+}^\mu(X,\tau) \right],
\end{equation}
where $\hat{a}^\mu = a^\mu + \Gamma^\mu_{\rho\sigma} a^{\rho\sigma}$ is the covariant form of $a^\mu$ and $F^\mu$ is a first order vector.
\par 

There exists another representation of the stochastic Newton equation that is given by
\begin{equation}
	F^\mu(X,\tau) =\frac{1}{2} m\left( D_+ D_- + D_- D_+ \right) X^\mu,
\end{equation}
where the \textit{covariant diffusion operators} $D_\pm$ act on an arbitrary first order $(k,l)$-tensor field $A(X,\tau)$ as, cf. Refs.~~\cite{DohrnGuerraI,DohrnGuerraII,Nelson},
\begin{equation}\label{eq:DiffusionOp}
	D_\pm A 
	= \left[
	\frac{\p}{\p \tau} 
	+ \hat{v}_\pm^\mu \nabla_\mu 
	+ \hat{v}_\pm^{\mu\nu} \left(\nabla_\mu \nabla_\nu + \mathcal{R}^{\,\cdot\,}_{\;\;\mu\,\cdot\,\nu} \right) \right] A,
\end{equation}
where
\begin{equation}
	\mathcal{R}^{\,\cdot\,}_{\;\;\alpha\,\cdot\,\beta} A^{\mu_1...\mu_k}_{\nu_1...\nu_l}
	=
	\sum_{i=1}^k \mathcal{R}^{\mu_i}_{\;\;\alpha\lambda\beta} A^{\mu_1 ... \mu_{i-1} \lambda \mu_{i+1} ... \mu_k}_{\nu_1...\nu_l}
	- \sum_{j=1}^l \mathcal{R}^{\lambda}_{\;\;\alpha\nu_j\beta} A^{\mu_1 ... \mu_k}_{\nu_1 ... \nu_{j-1} \lambda \mu_{j+1} ... \nu_l}.
\end{equation}
Using that $v_\pm^{\mu\nu} = \pm \frac{\hbar}{2m} g^{\mu\nu}$, eq.~\eqref{eq:DiffusionOp} can be rewritten as
\begin{equation}
	D_\pm A 
	= \left(
	\frac{\p}{\p \tau} 
	+ \hat{v}_\pm^\mu \nabla_\mu 
	\pm \frac{\hbar}{2m} \Box_{\rm DG} \right) A,
\end{equation}
where the \textit{Dohrn-Guerra Laplacian} is defined by
\begin{equation}
	\Box_{\rm DG} := g^{\mu\nu}\left(\nabla_\mu \nabla_\nu + \mathcal{R}^{\,\cdot\,}_{\;\;\mu\,\cdot\,\nu} \right).
\end{equation}

\subsection{Schr\"odinger equation}\label{sec:Schrodinger}
The solutions of the stochastic differential equation \eqref{eq:EQM} are stochastic processes. One can associate a probability density to these stochastic processes, and derive a partial differential equation for the evolution of this probability density. As argued in the introduction, the equation governing this evolution is the Schr\"odinger equation. Here, we present an explicit derivation.
\par 

Using eqs.~\eqref{eq:NablaS} and \eqref{eq:NablaR}, we can rewrite eqs.~\eqref{eq:HamJac23} and \eqref{eq:Continuity3} as
\begin{align}
	\frac{\p}{\p \tau} S(x,\tau) 
	&= 
	- \frac{1}{2m} \left(
	\nabla_\mu S \nabla^\mu S
	- \nabla_\mu R \, \nabla^\mu R 
	- \hbar \Box R  
	+ 2 \hbar A^\mu \nabla_\mu S
	+ \hbar^2 A_\mu A^\mu
	+ \frac{\hbar^2}{6} \mathcal{R} \right) 
	- \mathfrak{U},\\
	\frac{\p}{\p \tau} R(x,\tau) 
	&= 
	- \frac{1}{m} \left(
	\nabla_\mu S\, \nabla^\mu R
	+ A^\mu \nabla_\mu R
	+ \frac{\hbar}{2}\Box S
	+ \frac{\hbar^2}{2} \nabla_\mu A^\mu\right).
\end{align}
If we define the wave function
\begin{equation}
	\Psi(x,\tau) = e^{\frac{1}{\hbar}(R + i S)},
\end{equation}
we find that these equations are equivalent to the equation
\begin{equation}\label{eq:Schrodinger}
	i \hbar \frac{\p}{\p\tau}\Psi 
	= \left\{ - \frac{\hbar^2}{2m}\left[\left(\nabla_\mu + i A_\mu\right) \left(\nabla^\mu + i A^\mu\right) - \frac{1}{6} \mathcal{R} \right] + \mathfrak{U} \right\} \Psi.
\end{equation}
This is a generalization of the Schr\"odinger equation to pseudo-Riemannian geometry.\footnote{Note that for flat space-times $\mathcal{R}=0$. Moreover, in the non-relativistic limit one replaces $x^\mu\rightarrow x^i$ and identifies $\tau=t$. Therefore, in the flat non-relativistic limit we obtain the standard Schr\"odinger equation.} We note that the Born rule is an immediate consequence:
\begin{equation}
	|\Psi(x,\tau)|^2 = e^{\frac{2}{\hbar} R(x,\tau)} = \rho(x,\tau)
\end{equation}
by the definition of $R$ in eq.~\eqref{eq:Rdef}.

\subsection{Conformal coupling}\label{sec:Conformal}
In this section, we show that the generalization of the Schr\"odinger equation \eqref{eq:Schrodinger} imposes a conformal coupling of massive scalar particles to gravity. For this, we consider the Lagrangian of a free scalar field non-minimally coupled to gravity
\begin{equation}
	\mathcal{L}(\phi,\nabla\phi) 
	= 
	- \frac{1}{2} \left( \nabla_\mu \phi\, \nabla^\mu \phi + \frac{m^2}{\hbar^2}\, \phi^2 +	\xi \, \mathcal{R}\, \phi^2 \right).
\end{equation}
The field equation is given by the Klein-Gordon equation 
\begin{equation}
	\Box \, \phi = \frac{m^2}{\hbar^2}\, \phi + \xi\, \mathcal{R}\, \phi.
\end{equation}
We can construct an explicitly proper time dependent field $\Phi$ defined on $\mathcal{M}\times T$, such that
\begin{equation}
	\Phi(x,\tau) = \phi(x)\, e^{\frac{i m}{2\hbar}\tau},
\end{equation}
where $x=(t,\vec{x})$ is a four-vector. Then $\Phi$ satisfies the generalized Schr\"odinger equation \eqref{eq:Schrodinger} with $A_\mu=0$, $\mathfrak{U}=0$ and conformal coupling $\xi=\frac{1}{6}$. This result can be generalized in a straightforward manner to the cases $A_\mu\neq0$ and $\mathfrak{U}\neq0$.
\par

We conclude that stochastic quantization predicts that any scalar test particle must be conformally coupled to gravity. It is expected that this result can be generalized to arbitrary scalar fields. However, proof of this latter statement can only be achieved within a field theory description of stochastic quantization.

\section{Discussion}
In this paper, we have reviewed some aspects of second order geometry and stochastic quantization, and shown that the combination of the two leads to a consistent quantum theory on manifolds. In addition, we have further developed second order geometry, and constructed the notion of a Lie derivative in this framework. Furthermore, we have provided new results within stochastic quantization. In particular, we have shown that a diffeomorphism invariant framework of stochastic quantization imposes a conformal coupling of massive spin-0 test particles. It is expected that this result can be generalized to arbitrary scalar fields, but a proof of such a generalization requires further study of a field theory framework.
\par 

Since stochastic quantization can be formulated on (pseudo-)Riemannian manifolds, it is a natural approach to explore quantum gravity. However, in order to do so, a major hurdle must still be overcome, which is a consistent extension to both bosonic and fermionic field theories. Until now only a few specific bosonic examples have been studied in this framework, see for example Refs.~\cite{Guerra:1973ck,GuerraRuggiero,Guerra:1980sa,Guerra:1981ie,Kodama:2014dba,Marra:1989bi,Morato:1995ty,Garbaczewski:1995fr,Pavon:2001}, but no general formalism has yet been developed. The embedding of stochastic quantization into second order geometry, as developed in this paper could help guide the way towards such an extension.  Particularly interesting in this respect are recent developments in the study of Lagrangian dynamics on higher order jet bundles, see e.g. Refs.~\cite{Campos:2009ue,Campos:2010ay}, as this is the natural extension of second order geometry to a field theory setting.
\par

There are several studies that can be performed within the stochastic quantization framework without going to a field theory description or to dynamical backgrounds. The stochastic differential equation \eqref{eq:EQM} allows to solve and simulate the motion of quantum mechanical spin-0 test particles charged under scalar and vector potentials in any geometry. Such a study would be particularly interesting when performed in black hole geometries. One can then calculate the probability that a particle hits the singularity\footnote{In stochastic quantization, geodesic incompleteness of the space-time does not imply that the particle ends up at the singularity. One should study the Brownian completeness of the geometry instead, see e.g. Section 5 in Ref.~\cite{Emery}.} or escapes the black hole. Furthermore, one can calculate the expected proper time until one of these events occurs. Also, higher moments such as the variance for these events can be calculated. Such calculations could provide microscopic insights into Hawking radiation and black hole thermodynamics.
\par

In this paper, we have restricted ourselves to time-like processes with positive mass. A formulation for space-like processes can be obtained by considering imaginary masses and by replacing the proper time with the proper distance. However, a theory for massless particles on null-like surfaces is not easily obtained from the theory presented in this paper, and deserves further study.
\par

There are many other issues that deserve further exploration within the stochastic framework. For example, as discussed in the introduction, there is no consensus yet on the resolution of Wallstrom's criticism. Moreover, the notion of spin in stochastic quantization is only partially understood, see e.g. Refs.~\cite{Dankel,Nelson,Fritsche:2009xu}. In this paper, we have focused on scalar particles, in the presence of commuting spin-0 and spin-1 fields and gravity. Extensions to fermions, non-commuting potentials and higher spin fields would be interesting to investigate.
\par 

Furthermore, the formulation of stochastic quantization presented here was entirely in a position representation. Investigation of the dual picture in terms of momenta deserves further exploration. Early considerations along these lines can for example be found in Ref.~\cite{Shucker}.
\par 

Another open question is whether stochastic quantization can be formulated on complex manifolds instead of real manifolds. An argument for such a construction is that the wave function resembles the probability density of a complex random variable $Z=X+iY$ with $dZ=(V+iU)d\tau$. Discussions along these lines can also be found in Ref.~\cite{Pavon:1995}. Related to this is the question whether the function $R$ can be interpreted as an action for the background field in a Wick rotated version of the theory. The action $S$ would then be related to the probability density for the coordinates $Y$.
\par 

Finally, the presence of an osmotic velocity in stochastic quantization could provide new insights in the nature of dark matter. In this respect, it is worth noticing that the kinetic energy in stochastic quantization does not only contain the classical kinetic energy given by $\frac{m}{2} g_{\mu\nu} v^\mu v^\nu$, but also the osmotic energy of the background field given by $\frac{m}{2} g_{\mu\nu} \hat{u}^\mu \hat{u}^\nu$. It is expected that the notion of osmotic energy is also present in a field theoretical extension of stochastic quantization. In such an extension it will take the shape of the kinetic term of additional fields that only interact gravitationally with other fields. This suggests that the osmotic energy could be interpreted as dark matter.
\par

We conclude that stochastic quantization is an interesting framework, that deserves further exploration. We are currently investigating several aspects of the theory along the lines mentioned above, and hope to report on it elsewhere.

\section*{Acknowledgments}
This work is supported by a doctoral studentship of the Science and Technology Facilities Council. I would like to thank Joshua Erlich for interesting discussions on stochastic quantization. Furthermore, I would like to thank Xavier Calmet for helpful comments on the manuscript.
\end{document}